%% file: ms.tex
\pdfoutput=1
\RequirePackage[english=usenglish]{hyphsubst}	
\RequirePackage{amsmath}						
\RequirePackage{fix-cm}							
\documentclass{article}
\usepackage[utf8]{inputenc}		
\usepackage[T1]{fontenc}		
\usepackage[T1]{url} 			
\usepackage[numbers]{natbib}	
\usepackage[USenglish]{babel} 	
\usepackage{txfonts}			
\usepackage{helvet}         	
\usepackage{courier}        	
\usepackage{makeidx}         	
\makeindex             			
\usepackage[pdftex,fixpdftex,hyperref,svgnames,rgb,showerrors]{xcolor}	
\usepackage[pdftex]{graphicx}  	
\usepackage{multicol}        	
\usepackage[bottom]{footmisc}	
\usepackage{microtype}			
\usepackage{xspace}				
\usepackage[xspace]{ellipsis}	
\usepackage{listings}			
\usepackage{flafter}			
\usepackage[printonlyused]{acronym}	
\usepackage{booktabs}			
\usepackage{siunitx}			
\usepackage{todonotes}			
\usepackage{titling}
\usepackage{caption}
\usepackage[affil-it]{authblk}

\usepackage[%
    hidelinks,					%
    breaklinks=true,            %
    plainpages=false,           %
    pdfpagelabels=true,         %
    bookmarks=false,            %
	hyperfootnotes=false,		%
	pdftex
]{hyperref}
\let\orgautoref\autoref
\providecommand{\Autoref}[1]
{%
\def\pseudopartsautorefname{Part}
\def\pseudochaptersautorefname{Chapter}
\def\pseudosectionsautorefname{Section}
\def\partautorefname{Part}%
\def\chapterautorefname{Chapter}%
\def\sectionautorefname{Section}%
\def\subsectionautorefname{Section}%
\def\subsubsectionautorefname{Section}%
\def\equationautorefname{Equation}%
\def\figureautorefname{Figure}%
\def\subfigureautorefname{Figure}%
\def\tableautorefname{Table}%
\def\lstlistingautorefname{Listing}%
\def\lstnumberautorefname{Line}%
\orgautoref{#1}%
}%
\renewcommand{\autoref}[1]
{%
\def\pseudopartsautorefname{Part}
\def\pseudochaptersautorefname{Chap.}
\def\pseudosectionsautorefname{Sect.}
\def\partautorefname{Part}%
\def\chapterautorefname{Chap.}%
\def\sectionautorefname{Sect.}%
\def\subsectionautorefname{Sect.}%
\def\subsubsectionautorefname{Sect.}%
\def\equationautorefname{Eq.}%
\def\figureautorefname{Fig.}%
\def\subfigureautorefname{Fig.}%
\def\tableautorefname{Table}%
\def\lstlistingautorefname{Listing}%
\def\lstnumberautorefname{line}%
\orgautoref{#1}%
}%
\definecolor{basiccolor}{rgb}{0, 0, 0}
\definecolor{stringcolor}{rgb}{0.7, 0.1, 0.2}
\definecolor{identifiercolor}{rgb}{0, 0, 0.8}
\definecolor{commentcolor}{rgb}{0.5, 0.5, 0.5}
\definecolor{keywordcolor}{rgb}{0, 0, 0.8}
\definecolor{emphcolor}{rgb}{1, 0, 0}
\lstdefinelanguage{ANDL}{
	morekeywords={inline, ini, network, devices, node, switch, gateway, canLink, communication, message, sender, receivers, payload, period, connections, segment, new, mapping, tt, avb, ctID, can, id, pool, holdUp, types, ethernetLink, bandwidth},
	sensitive=false,
	morecomment=[l]{\%},
	morecomment=[l]{//},
	morestring=[b]",
	identifierstyle=\color{basiccolor},
	keywordstyle=\color{keywordcolor}\bfseries,
	commentstyle=\color{commentcolor}\itshape,
	tabsize=2
}
\lstdefinelanguage{XMLconstraint}{
	morekeywords={constraints, constraint, module, name, min, max, avg_min, avg_max, moduleIsRegex, nameIsRegex, samples},
	sensitive=false,
	morecomment=[l]{\%},
	morecomment=[l]{//},
	morestring=[b]",
	moredelim=[s][\color{keywordcolor}]{>}{<},
	identifierstyle=\color{basiccolor},
	keywordstyle=\color{keywordcolor}\bfseries,
	commentstyle=\color{commentcolor}\itshape,
	tabsize=2
}
\lstdefinelanguage{OMNETPPINI}{%
    language=,				
	morekeywords=[1]{network,include,extends},	
	otherkeywords={sim-time-limit,num-rngs},	
	morekeywords=[2]{true,false,xmldoc},		
	morekeywords=[3]{uniform,exponential,normal},	
	identifierstyle=\color{Black}, 			
	stringstyle=\color{SeaGreen},			
	commentstyle=\itshape\color{Black!65}, 	
	keywordstyle=[1]\color{DarkBlue},		
	keywordstyle=[2]\bfseries\color{Purple},
	keywordstyle=[3]\color{Purple},			
	morestring=[b]",		
	morecomment=[l]{\#},	
	sensitive=true 		
}
\lstdefinelanguage{OMNETPPNED} {
    morekeywords=[1]{parameters,gates,submodules,connections,allowunconnected},
    morekeywords=[2]{bool,channel,channelinterface,const,
                     default,double,extends,false,for,if,import,index,inout,input,
                     int,like,module,moduleinterface,network,output,package,
                     property,simple,sizeof,string,this,true,types,volatile,
                     xml,xmldoc},
	keywordstyle=[1]\color{DarkBlue},		
	keywordstyle=[2]\bfseries\color{Purple},
    sensitive=true,
    morecomment=[l]{//},
    morestring=[b]",
}
\lstdefinelanguage{OMNETPPMSG} {
    morekeywords={abstract,bool,char,class,cplusplus,double,enum,extends,false,
                  fields,import,int,long,message,namespace,noncobject,packet,
                  properties,readonly,short,string,struct,true,unsigned},
    sensitive=true,
    morecomment=[l]{//},
    morestring=[b]",
}
\DeclareRobustCommand*{\omnet}{{OMNeT\nolinebreak\hspace{-0.025em}\texttt{++}}\xspace}	
\newcommand{\code}[1]{\texttt{#1}\xspace}
\newcommand{\cursive}[1]{\textit{#1}\xspace}
\newcommand{\program}[1]{\textbf{#1}\xspace}
\newcommand{\filename}[1]{\textsl{#1}\xspace}
\usepackage{xparse}		
\DeclareDocumentCommand\Code{ v }{%
    {\texttt{#1}\xspace}%
}%
\DeclareDocumentCommand\Cursive{ v }{%
    {\textit{#1}\xspace}%
}%
\DeclareDocumentCommand\Program{ v }{%
    {\textbf{#1}\xspace}%
}%
\DeclareDocumentCommand\Filename{ v }{%
    {\textsl{#1}\xspace}%
}%
\DeclareDocumentCommand\keys{ m g g }{%
    {[\textit{#1}]%
        \IfValueT {#2} {\texttt{+}[\textit{#2}]}%
		\IfValueT {#3} {\texttt{+}[\textit{#3}]}%
     \xspace%
	}%
}%
\DeclareDocumentCommand\commands{ m g g g g }{%
    {\textit{#1}%
        \IfValueT {#2} {\hspace{0.15ex}$\rightarrow$\hspace{0.15ex}\textit{#2}}%
		\IfValueT {#3} {\hspace{0.15ex}$\rightarrow$\hspace{0.15ex}\textit{#3}}%
		\IfValueT {#4} {\hspace{0.15ex}$\rightarrow$\hspace{0.15ex}\textit{#4}}%
		\IfValueT {#5} {\hspace{0.15ex}$\rightarrow$\hspace{0.15ex}\textit{#5}}%
     \xspace%
	}%
}%
\graphicspath{{figures/}}		
\begin{document}				
%
\input{book-structure}		
%
%
\title{Simulation of Mixed Critical\\ In-vehicular Networks}
\author[1]{Philipp Meyer\thanks{Corresponding author: philipp.meyer@haw-hamburg.de}}
\author[1]{Franz Korf}
\author[2]{Till Steinbach}
\author[1]{Thomas C. Schmidt}
\affil[1]{\small{Hamburg University of Applied Sciences

Department Informatik

Hamburg, Germany}}

\affil[2]{\small{Ibeo Automotive Systems GmbH

Hamburg, Germany}}
\date{\today}
\maketitle
%
%
\newcommand{\abs}{%
Future automotive applications ranging from advanced driver assistance to autonomous driving will largely increase demands on in-vehicular networks. Data flows of high bandwidth or low latency requirements, but in particular many additional communication relations will introduce a new level of complexity to the in-car communication system. It is expected that future communication backbones which interconnect sensors and actuators with \acp{ECU} in cars will be built on Ethernet technologies. However, signalling from different application domains demands for network services of tailored attributes, including real-time transmission protocols as defined in the \ac{TSN} Ethernet extensions. These \ac{QoS} constraints will increase network complexity even further. Event-based simulation is a key technology to master the challenges of an in-car network design. This chapter introduces the domain-specific aspects and simulation models for in-vehicular networks and presents an overview of the car-centric network design process. Starting from a domain specific description language, we cover the corresponding simulation models with their workflows and apply our approach to a related case study for an in-car network of a premium car.
}%
\begin{abstract}
  \abs
\end{abstract}
\acresetall		
%
%
\include{introduction}
\include{mixed_critical}
\include{simulation_environment}
\include{simulation_process}
\include{case_study}
\include{conclusion}
%
%
\input{acronyms}
%
%
\bibliographystyle{dinat}
\bibliography{references}
\end{document}

%% file: book-structure.tex
%
%
%
%
\newcounter{pseudoparts}
\renewcommand*\thepseudoparts{\textit{\{\Roman{pseudoparts}\}}}
\newcounter{pseudochapters}
\renewcommand*\thepseudochapters{\textit{\{\arabic{pseudochapters}\}}}
\newcounter{pseudosections}
\renewcommand*\thepseudosections{\textit{\{\arabic{pseudochapters}.\arabic{pseudosections}\}}}
%
\refstepcounter{pseudoparts}\label{part:omnet-simulator}
	%
	%
	\refstepcounter{pseudochapters}\label{chapter:omnet}
	%
\refstepcounter{pseudoparts}\label{part:omnet-ecosystem}
	%
	%
	\refstepcounter{pseudochapters}\label{chapter:inet}
	%
	%
	\refstepcounter{pseudochapters}\label{chapter:inetmanet}
	%
	%
	\refstepcounter{pseudochapters}\label{chapter:rinasim}
	%
	%
	\refstepcounter{pseudochapters}\label{chapter:simulte}
	%
	%
	\refstepcounter{pseudochapters}\label{chapter:veins}   
	%
	%
	\refstepcounter{pseudochapters}\label{chapter:sea}    
	%
\refstepcounter{pseudoparts}\label{part:omnet-research}
	%
	%
	\refstepcounter{pseudochapters}\label{chapter:reproducibility}
	%
	%
	\refstepcounter{pseudochapters}\label{chapter:remotecontrol}
	%
	%
	\refstepcounter{pseudochapters}\label{chapter:invehicular}   
	%
	%
	\refstepcounter{pseudochapters}\label{chapter:limosim}
	%
	%
	\refstepcounter{pseudochapters}\label{chapter:arteryvanetza}
	%
	%
	\refstepcounter{pseudochapters}\label{chapter:ltevehicular}
    %
	%
	\refstepcounter{pseudochapters}\label{chapter:oppnets}
	%
	%
	\refstepcounter{pseudochapters}\label{chapter:opendsme}
	%

%% file: introduction.tex
\section{Introduction}
\label{sec:introduction}
The automotive market is growing in demand for innovative driver assistance systems, as well as highly automated or even autonomous driving units.
In-vehicular communication networks that connect sensors and actuators with \acp{ECU} contribute the basis to these distributed, safety-critical, and highly complex systems. Consequently, their architecture and design are playing an increasingly important role.
As of today, in-car communication concepts fall short in meeting the emerging requirements of future driving systems.

High bandwidth demands from distributed visual sensors---the raw data fusion of laser scanners and cameras for example---exceed the capacities of current data transmission systems by more than an order of magnitude. For example a low resolution camera stream of \SI{7}{\mega\bit\per\second} already exceeds CAN's \SI{0.5}{\mega\bit\per\second} 14 times.
An increasing number of vehicular safety functions pose strict redundancy or quality of service  requirements such as latency and jitter.
 With respect to this growing heterogeneity, current automotive communication architectures and technologies reach their limits.
With timing and bandwidth aspects in mind, communication techniques that provide a wide range of real-time communication services are needed.
Due to its high data capacities, its low cost of commodity components, and its large flexibility in terms of protocols and topologies, switched Ethernet is a promising candidate to overcome the challenges of future in-car networks \cite{mk-ae-15}.

Communication architectures of today's vehicles are composed of different domain-specific technologies such as \ac{CAN}, FlexRay, \ac{LIN}, and \ac{MOST} \cite{rb-baeae-13}. Cross-domain communication is enabled via a central gateway that inter-connects a majority of these buses.
For premium cars, the simple structuring mechanism of a central gateway reaches its limits in terms of complexity and controllability.
Future developments of  automotive services and communication require new concepts and solutions.

The \ac{OPEN} Alliance Special Interest Group, which is driven by the automobile industry, focuses on the standardization of certified automotive Ethernet that runs over one single pair of unshielded twisted wires, previously offered as BroadR-Reach by Broadcom \cite{ieee8023bw,ieee8023bp}.
It is very well suited for the challenging \ac{EMC} requirements compliant to the harsh environment in cars.
This standard lays the foundation of automotive Ethernet variants.
Since established automotive suppliers already offer this technology, Ethernet is a candidate for a new common communication architecture in vehicles \cite{hrbs-ctets-07,sks-ctefe-10,slksh-tiice-12}.

When switching to Ethernet the in-vehicle network will face a significant paradigmatic change. However, sudden changes in the network architecture of mass-produced cars is infeasible due to costs and risks.
That is why there will be a gradual transition to a flat Ethernet topology.
First steps involve the migration of Ethernet into legacy bus topologies.
A consolidation strategy with heterogeneous networks formed of an Ethernet core and legacy buses at the edges will allow to preserve investments in knowledge about legacy technologies.
Such a mixed architecture forms the beginning of this stepwise transition towards a flat network topology consisting solely of Ethernet links \cite{msks-reppa-11}.

The design of future vehicular networks requires new tools for experimentation, optimization, and debugging.
There are several commercial tools to analyze in-car networks.
In industry, most popular is \program{CANoe} (by Vector Informatik GmbH) that enables real-time cluster simulations of fieldbuses.
As of today, \program{CANoe} does not provide functionality to simulate real-time Ethernet variants.
\program{SymTA/S} is a commercial timing analyzer (not a network simulator) by Symtavision GmbH that supports Ethernet (standard and \ac{AVB} \cite{ieee8021ba-11}) as well as common fieldbus technologies.
It provides analytical models to calculate load and timing.

To evaluate future in-vehicle networks, appropriate tools need to integrate the distributed system components.
While current tool chains focus on bit-correct simulation of fieldbus communication, future environments have to enable the developer to analyze effects of congestion and jitter on the cars applications and assistance functions on a system level.
For example it can easily be explored how a third message source is influencing the traffic between two communication partners.
System level simulation increases the understanding of the behavior of future automotive communication architectures and enable quantitative analysis at a level of higher abstraction.

The \program{\omnet} (see \autoref{part:omnet-simulator}) platform is a well suited tool and a perfect base to implement such kind of system level simulation based on real-time Ethernet variants \cite{sdks-eoifs-11}.
Besides its open-source simulation core, it allows to extend its Eclipse based IDE with custom plugins for specialized design and analysis tasks.
In this work, we introduce both a uniform workflow as well as the required models and tools to design and evaluate future in-vehicle networks.

Experiences with the simulation during research on in-car network architectures showed that the configuration of these large networks is complex and tedious. Thus there was a demand to simplify the description of in-car network scenarios. This demand led to the development of a \ac{DSL} that supports the fast setup of simulations of in-car network architectures.

This DSL is called \program{\ac{ANDL}} and is integrated as a plugin for \program{\omnet}. It enables the design of networks on an abstract layer. Additionally benefits like autocompletion, syntax highlighting, validation, autoformatting, renaming and scoping support the configuration process.

The remainder of this chapter is structured as follows. In \autoref{sec:mixed_critical}, we discuss the problem space of mixed critical networks in cars. Our simulation environment including  tools and models is introduced in \autoref{sec:simulation_environment}. \Autoref{sec:simulation_process} shows an example of our simulation workflow, followed by a case study about backbones in premium cars in \autoref{sec:case_study}. We conclude with an outlook in \autoref{sec:conclusion}.

%% file: mixed_critical.tex
\section{Mixed Critical In-vehicle Networks}
\label{sec:mixed_critical}
In a current premium car, there are up to 70 electronic control units (ECU) with more than 900 functions interconnected over a heterogeneous in-car network. While control loop applications have strong real-time requirements, other applications such as navigation, firmware updates or multimedia streaming demand high bandwidth at relaxed timing constraints. With the introduction of high quality sensors such as high-resolution driver assistance camera systems some functions require high date rates and  rigid timing. For safety critical functions like autonomous driving, timing and data rates must be strictly guaranteed. This leads to a wide spectrum of soft and hard time-constrained domains, some of which covering the entire topology.

The inter- and intra-domain communication is growing and the amount of data exchanged within the car is heavily increasing.  In addition to on-board systems, a car will receive off-board data by its backend or by other cars, and infrastructure components such as traffic lights  from the vicinity will use car-to-car (C2C) or car-to-infrastructure (C2I) communication.

\begin{figure}[!ht]
  \centering
  \includegraphics[width=1\textwidth]{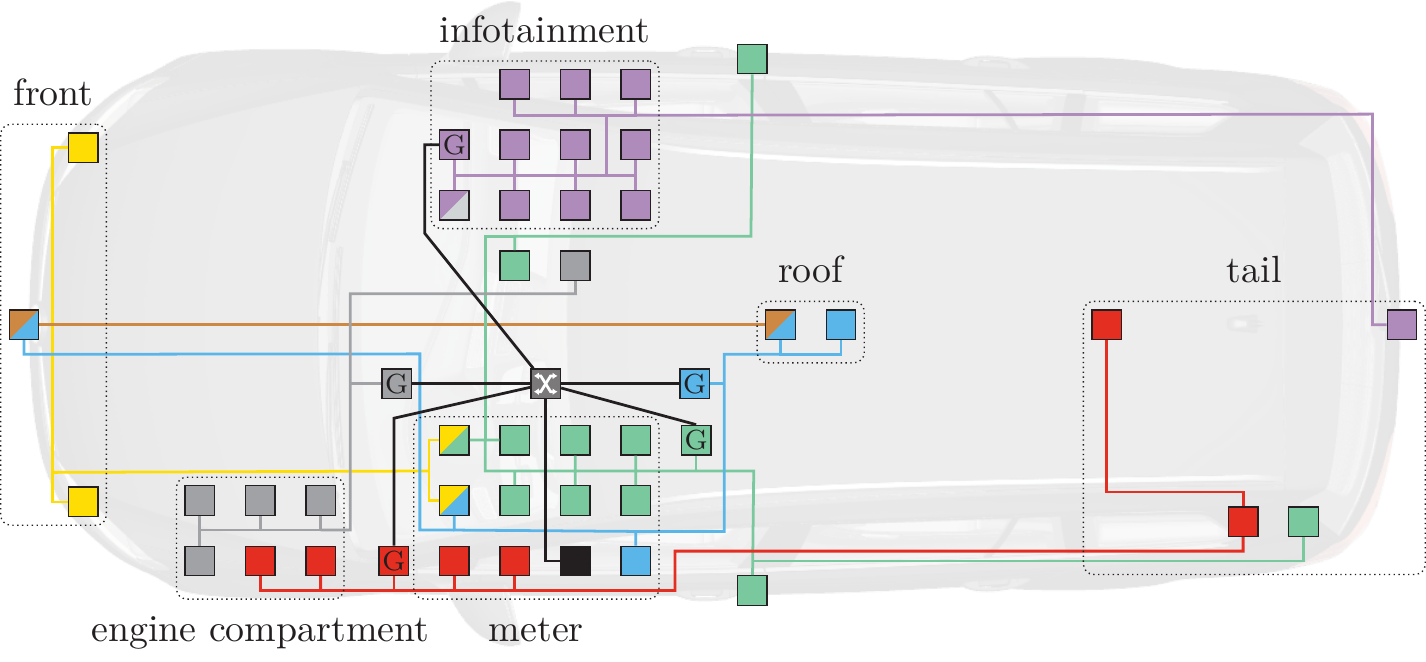}
  \caption{Domain Decomposition of a Traditional Car Network}
  \label{fig:domains}
\end{figure}

Ethernet is the key technology discussed by the major OEMs to overcome the challenges of future in-car networking \cite{mk-ae-15}. Consequently, the automobile industry is pushing standardization of a physical layer for automotive applications within the \ac{OPEN} Alliance Special Interest Group. The \SI{100}{\mega\bit\per\second} automotive certified physical layer is already available (commercially available as BroadR-Reach, standardization by the IEEE under P802.3bw \cite{ieee8023bw}), \SI{1}{\giga\bit\per\second} automotive links are standardized under 802.3bp \cite{ieee8023bp}.

One possible direction for building future automotive communication is a homogeneous core network of switched Ethernet. Such a flat design reduces complexity by purely switching without the need for gateways between different technological domains in the car. On the other hand, OEMs need to protect their investments in fully developed and proven ECUs as well as their software components. In most cases, these components follow an integrated design that communicates via CAN. Changing to Ethernet hardly justifies the redesign of these components. As part of the migration to a pure Ethernet-based communication layer, gateways will connect CAN buses to the Ethernet backbone (s. Figure \ref{fig:domains}). Corresponding gateways must support a tunneling of CAN message over the Ethernet backbone to interconnect  CAN buses of different domains. In addition, CAN message priorities must be preserved. Since CAN supports a maximum payload of \SI{8}{bytes} and Ethernet offers a minimum payload of \SI{46}{bytes}, these gateways will allow the aggregation of CAN messages within an Ethernet frame.

Thinking these directions of design it becomes evident that an in-car network can no longer be considered a collection of closed domains with fixed, offline-configured traffic. Instead, dynamic traffic and changing communication requirements must be foreseen in particular with the integration of new services entering the car via an Internet uplink. Examples for such applications include online software updates, car diagnoses, or updates of on-board information repositories such as navigation maps or meta-data.
Such networked applications are easy to host on an Ethernet-based in-car backbone. Even an offloading to a cloud of data or computationally expensive tasks are currently discussed in the context of autonomous driving.  Envisioned from these perspectives, future cars may even be characterized as mobile constituents  of the Internet of Things (IoT), requiring the network to cope with ever arriving new challenges that  include  security and safety.

\subsection{Time Sensitive Networking Technologies}

Automotive communication consists of a  collection of distinct services that strictly differentiate in quality, and potentially interfere within a flat common network \cite{slksh-bhcan-15}.
Hence, standard switched Ethernet must be extended beyond simple traffic prioritization to provide real-time guarantees. Typical techniques are bandwidth limitation as in IEEE 802.1Qav \cite{ieee-802.1qav}, or  rate-constraining as  in TTEthernet or AFDX \cite{arinc-664} traffic. Further time-triggering such as the time-triggered traffic class of AS6802 \cite{as6802-11} or the upcoming IEEE 802.1Qbv (Enhancements for Scheduled Traffic) \cite{ieee8021qbv-13}. The upcoming Time Sensitive Networking (TSN) standard bundles all of these techniques.

\begin{figure}[b]
\centering
\includegraphics[width=1\columnwidth]{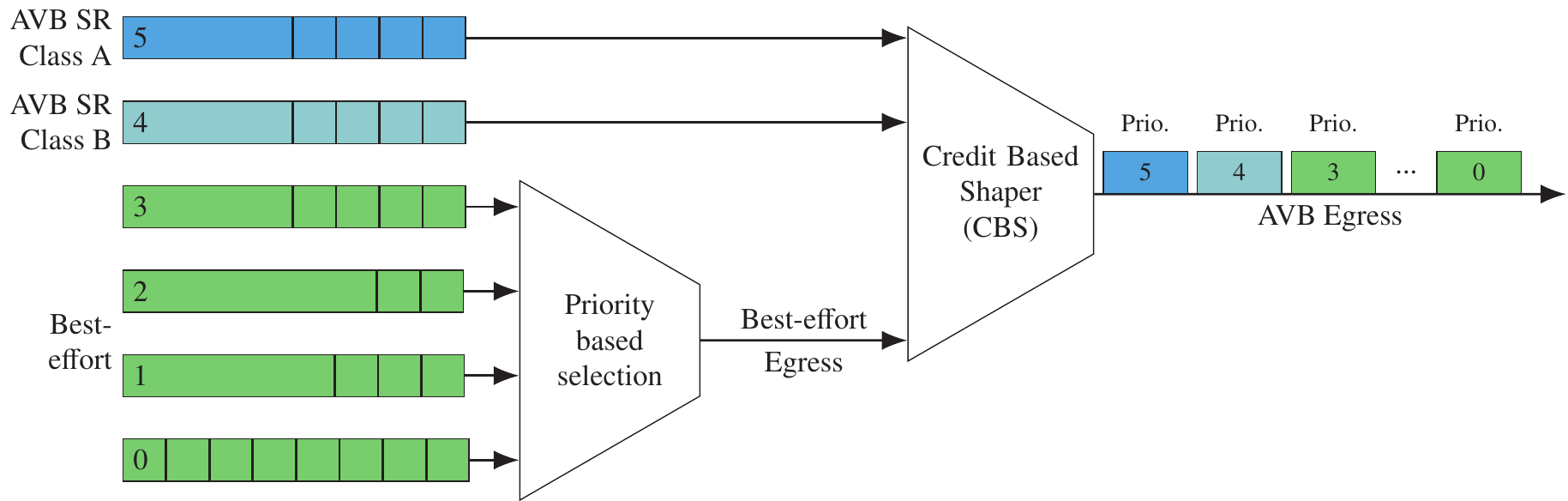}
\caption{IEEE 802.1Qav: Transmission Selection Algorithms}
\label{fig:scheduler}
\end{figure}

The IEEE 802.1 Audio/Video Bridging (AVB) standard \cite{ieee8021ba-11} is a core predecessor  of TSN. It enables low latency streaming services and guaranteed data transmission in switched Ethernet networks. This real-time Ethernet extension originates from the multimedia domain where synchronization, jitter and latency constraints of the applications are high. Ethernet AVB guarantees latencies under \SI{2}{\ms} over  seven hops for its best traffic class. The IEEE 802.1 AVB standard consists of different sub-standards required to guarantee the latency, synchronization performance, as well as a coexistence with legacy Ethernet nodes.

\emph{IEEE 802.1Qav} \cite{ieee-802.1qav} specifies queuing and forwarding rules to guarantee the latency constraints for AVB and the support of legacy Ethernet frames. AVB defines two service classes with different guarantees,

\begin{enumerate}
 \item Stream reservation class-A with a maximum latency of \SI{2}{\ms}
  \item Stream reservation class-B with \SI{50}{\ms} over seven hops.
\end{enumerate}
An AVB network is also able to deal with non-AVB frames. These frames are mapped to the best-effort class (see Fig. \ref{fig:scheduler}).

Prioritization, queuing and scheduling mechanisms realize a guaranteed data transmission of AVB frames within strict latency bounds. A transmission of an AVB frame is controlled by using a credit based shaper (CBS). Transmission of an AVB frame is allowed when the number of available credits is larger or equal 0. Implicitly, the CBS has a lower and upper bound to limit the data rate and burstiness of AVB data. The remaining bandwidth is available for non-AVB nodes. To ensure that AVB traffic always has the highest priority, the priority of legacy Ethernet frames by non-AVB nodes is re-mapped to the priorities of the best-effort traffic class. Furthermore, there is a signaling protocol specified in \emph{IEEE 802.1Qat} \cite{ieee8021qat-10} to reserve the required resources for AVB frames along the entire path between source and sink. The standard recommends that at most \SI{75}{\percent} of the total bandwidth shall be reserved for AVB data, while the remaining resources are freely available to best-effort traffic.

Another option of traffic shaping and media access policy for real-time communication in switched networks is time-triggered Ethernet. The TTEthernet protocol was standardized in 2011 by the Society of Automotive Engineers (SAE) \cite{saeas2d} under AS6802 \cite{as6802-11}. It is a compatible extension of IEEE switched Ethernet and uses topologies formed of full-duplex links. The TTEthernet media access strategies are similar to IEEE 802.1Qbv (Enhancements for Scheduled Traffic) that is under development by the IEEE Time Sensitive Networking (TSN).

Time-triggered Ethernet variants are operating on an offline configured schedule with dedicated transmission slots for all real-time messages shared among all network participants. This enables a \emph{coordinated} time-division-multiple-access (TDMA) media access strategy with deterministic transmission and predictable delays. TDMA prevents congestion on outgoing line cards and thereby enables isochronous communication with low latency and jitter. To allow for this access scheme, a failsafe synchronization protocol has to provide a precise global time among all participants.

In addition to time-triggered, TTEthernet defines two other event-triggered message classes: \emph{Rate-constrained} (RC) is comparable to the link layer of the \emph{ARINC-664 (AFDX)} protocol \cite{arinc-664}. Bandwidth limits for each stream and sender enable the real-time guarantees. So-called bandwidth allocation gaps (BAGs) implement the bandwidth limits. The BAGs define the minimum distance of two consecutive frames of the same stream (called virtual link). The rate-constrained traffic is comparable with Ethernet AVBs stream reservation classes A and B. Similarly it uses strict priorities for traffic with different real-time requirements.

The \emph{Best-effort} (BE) traffic conforms to standard Ethernet messages transmitted with the lowest priority. The best-effort class is used for the transmission of cross-traffic. It allows the integration of hosts that are unaware of the time-triggered protocol and remain unsynchronized.

%% file: simulation_environment.tex
\section{Simulation Environment}
\label{sec:simulation_environment}

\begin{figure}[h]
\includegraphics[width=\linewidth]{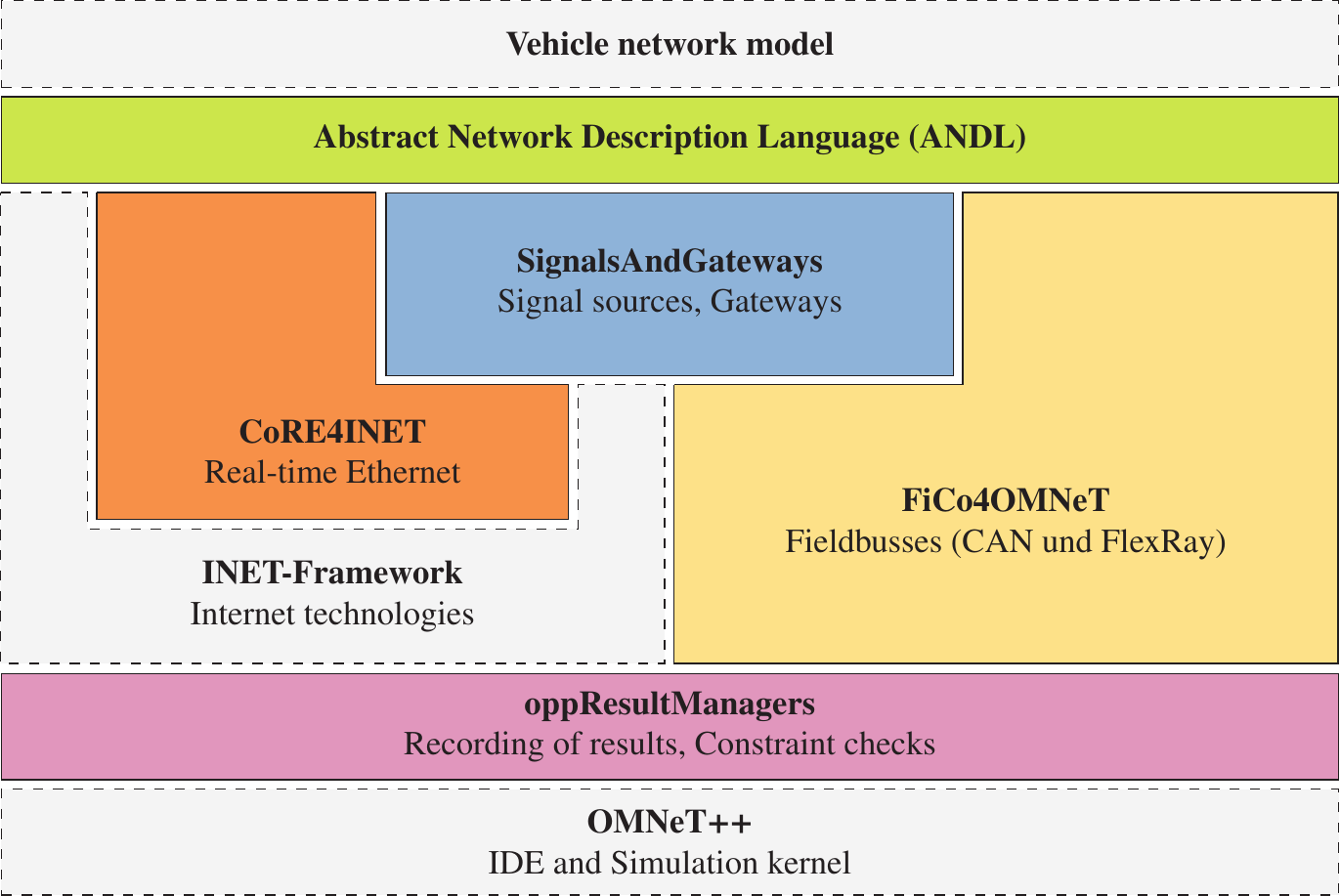}
\caption{Overview of the contributed simulation environment (colored)}
\label{fig:models}
\end{figure}

The simulation models introduced in this section were developed for simulation of in-car networks but can be used for other systems as well. To simplify the installation, an \program{\omnet} plug-in is provided that offers an automated installation process as well as an update procedure. \Autoref{fig:models} gives an overview of the contributed simulation models and their place in the software stack of the toolchain.

All simulation models base on the IDE and simulation kernel of \program{\omnet}. The optional \program{oppResultManagers} framework writes simulation results directly to PCAPng files or a database. \program{FiCo4OMNeT} contributes simulation models for CAN and FlexRay fieldbuses. For both there are no dependencies to other simulation models.
\program{CoRE4INET} provides simulation models for real-time Ethernet communication. It uses the Ethernet layer implemented in the \program{INET} framework. \program{INET} models of higher layer can communicate via \program{CoRE4INET} models, too.
\program{SignalsAndGateways} implements gateways. They support different strategies for the translation of communication between real-time Ethernet and fieldbuses. Thus, it depends on \program{CoRE4INET} and \program{FiCo4OMNeT}.
\\
As simulation input a vehicle network must be described using  \filename{*.ini} and \filename{*.ned} files of \program{\omnet}.
It is time consuming to describe several variants of a vehicle network using these files.
Hence, such a network should be modelled in the domain specific language \program{ANDL}.
A compiler translates network designs given in \program{ANDL} to the corresponding \filename{*.ini} and \filename{*.ned} files used by \program{CoRE4INET}, \program{FiCo4OMNeT} and \program{SignalsAndGateways} models.
\\
All simulation models and plug-ins are published open-source and can be downloaded from our website \cite{core-frameworks}.

\subsection{Domain Specific Language for Automotive Networks}
\label{subsec:dsl}
Configuring the simulation of large heterogeneous networks is complex and lengthy. To reduce this effort and to let the developer focus on the design task, we developed a \ac{DSL} for the description of heterogeneous in-vehicle network designs. It is called \acl{ANDL} and provides an easy and assisted way to design a network in an Eclipse environment. It is implemented as an Eclipse plugin and thus fits into the \program{\omnet} IDE. The plugin provides syntax highlighting as well as context aware code completion.
For typical vehicle networks that require few TDMA based communications, a scheduling algorithm generates a feasible schedule for TDMA traffic. It should be used as starting point for improving the TDMA schedule \cite{ksks-dtina-14}.

The \program{ANDL} is implemented as an \program{\omnet} plugin using Eclipse's Xtext technology \cite{xtext, l-idslx-16}. Xtext is a framework for development of programming - and domain-specific languages.
Using a grammar that has been extended by some specific elements, the DSL will be described. Based on this input Xtext generates a parser and a code editor that will be plugged into \program{\omnet} IDE. Using a set of Java classes, which have been generated by Xtext, the compilation from \program{ANDL} to \filename{*.ini} and \filename{*.ned} of \program{\omnet} will be defined.

\subsection{CoRE4INET}
\label{subsec:core4inet}
\program{CoRE4INET} (Communication over Real-time Ethernet for \program{INET}) is a suite of real-time Ethernet simulation models. Currently it supports the AS6802 protocol suite, traffic shapers of Ethernet AVB, IEEE 802.1Q, and models for mapping IP traffic to real-time traffic classes.

The centre of the \program{CoRE4INET} models is the implementation of media access strategies for different traffic classes. By combining these strategies, new traffic shapers can be designed that are able to forward real-time traffic of different standards on the same physical link. For example it is possible to combine time-triggered traffic of AS6802 with credit based shaping of Ethernet AVB to form a new time-aware shaper that can handle both classes in parallel \cite{msks-eatts-13}. This allows to evaluate new concepts that are currently under standardization or are even not yet assessed.

For incoming traffic, the models contain traffic selection and constraint checks. To simulate time-triggered behaviour and time-synchronization, \program{CoRE4INET} provides models for oscillators, timers and schedulers. Oscillators allow to implement the behaviour of inaccurate clocks with their unique influence on real-time communication. Finally, \program{CoRE4INET} contains application models for simple traffic patterns and traffic bursts.
\\
Selected simulation models were checked against analytical models of the different specifications and evaluated in empirical tests using real-world hardware \cite{sks-ctefe-10}.

\subsection{FiCo4OMNeT}
\label{subsec:fico4omnet}
CAN \cite{wo-can-13} is a fieldbus widely used in automobiles. Future vehicle networks require mixed operation of Ethernet and CAN.
FlexRay \cite{eb-f-12} is used in a few premium vehicles. For the next generation of these vehicles a migration from FlexRay to real-time Ethernet is expected.
Hence, \program{FiCo4OMNeT} (Fieldbus Communication for \program{\omnet}) provides simulation models for CAN and FlexRay.

Exploiting the fact that all FlexRay nodes are connected to the same bus, the static segment of FlexRay provides a TDMA based communication.
FiCo4OMNeT implements this behavior using a clock and a oscillator model, which are simpler than the one of \program{CoRE4INET}.
In order for the two models to remain independent, \program{FiCo4OMNeT} provides a simple clock- and oscillator model.
Similar to \program{CoRE4INET}, it contains application models for CAN and FlexRay applications with simple traffic patterns.
The fieldbus models in \program{FiCo4OMNeT} were originally checked against results of the \program{CANoe} simulation environment \cite{bsks-stafc-13}, an industry standard software for the analysis of CAN bus communication.

\subsection{SignalsAndGateways}
\label{subsec:signalsandgateways}
Using gateways, the \program{SignalsAndGateways} simulation model interlinks between real-time Ethernet and fieldbusses. These gateways are specific network nodes that translate between legacy bus technologies and (real-time) Ethernet. To be as flexible as possible, a gateway consists of three submodules:

\paragraph*{\bf Path finding}
The router module decides to which components an information is forwarded.
It receives messages in their original representation (CAN - or Ethernet frame).
Based on forwarding rules it selects the path that the message will take.
Using the header information of the message, a forwarding rule defines a CAN bus or an Ethernet node that should receive the information of the message.
A static defined routing table stores all forwarding rules.
If there is no entry in the routing table for a message, it is dropped.
Otherwise it will be sent to all destination given by the forwarding rules that match to the message.
\\
There is no limit of busses and links a gateway can be connected to. The gateway can also translate between fieldbus technologies, thus it is also applicable to legacy designs with multiple busses interconnected over a central gateway.

\paragraph*{\bf Buffering}
Gateways support aggregation strategies to improve bandwidth utilization of different technologies. CAN messages for example have a maximum payload of \SI{8}{\byte}, while Ethernet messages have a minimum payload of \SI{46}{\byte}. If an Ethernet frame encapsulates only one CAN message, the rest of minimum payload would be padded and bandwidth would be wasted. Aggregation strategies implemented in the buffer modules allow to release frames in groups, according to different strategies. These strategies are implemented in the buffer modules, too.

Aggregation strategies have a huge impact on the latency of messages passing a gateway. All strategies delay frames to collect multiple messages before aggregating them into a single frame. The most popular strategy implemented in the buffer is the pooling strategy with holdup time. The holdup time of a message defines the maximum acceptable delay for this message.
Each message is assigned to a pool, while multiple different messages share the same pool. To each message a holdup time is assigned. On arrival of a frame in the pool, its holdup time is compared with the holdup time of the pool. If the holdup time of the frame is shorter, the holdup time of the pool is adjusted correspondingly. When the holdup time of the pool is expired all messages in the pool are released together in one frame.
The modular architecture of the gateway allows to easily add more aggregation strategies.

\paragraph*{\bf Transformation}
Transformation modules implement the translation between different communication technologies. The strategies transparently map information between fieldbuses and Ethernet. Currently there is a simple mapping between fieldbus frames and raw (layer 2) Ethernet frames. The modular architecture of the gateway allows to easily add more sophisticated mappings, e.g. when higher layer application protocols should be used.

Similar to real-world gateways, gateway nodes can host applications that are not related to gateway functionality. Thus gateways can be added to control units that also host application software.

\subsection{Result Manager}
\label{subsec:result_manager}
The \program{\omnet} IDE already comes with tools for the result analysis. We extended those built-in tools to simplify the analysis in specialized use-cases and developed interfaces to interconnect the \program{\omnet} simulation with established industry products.

\program{oppResultManagers} is a set of modules for \program{\omnet} simulations. Instead of simulation models it contains so called ResultManagers. ResultManagers are responsible for writing out simulation results. The \program{\omnet} vector and scalar files, as well as the eventlog are built-in instances of ResultManagers. The \program{oppResultManagers} project adds ResultManagers for example \program{PCAPng}, \program{SQLite}, \program{postgreSQL} and Constrained Check. It is also possible to use several managers in parallel.

%% file: simulation_process.tex
\section{Simulation Process}
\label{sec:simulation_process}
\begin{figure}[!ht]
  \centering
  \includegraphics[width=1\textwidth]{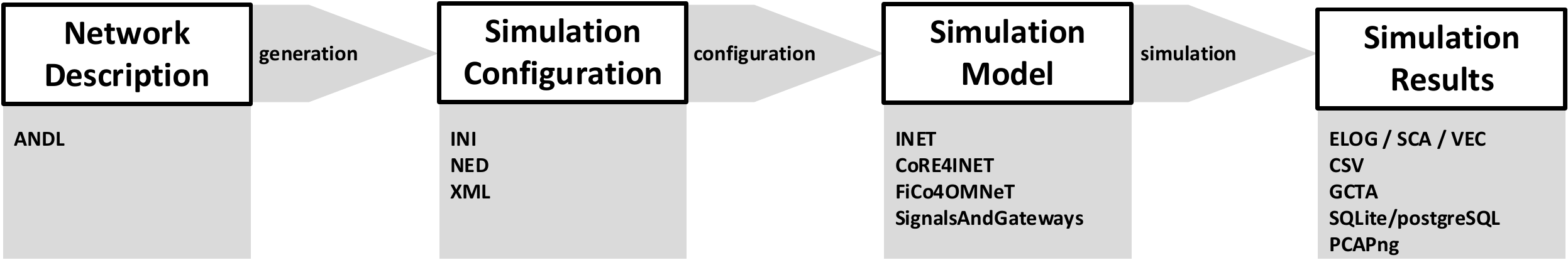}
  \caption{Workflow of simulation projects -- from network description to result analysis}
  \label{fig:workflow}
\end{figure}

The simulation process (s. figure \ref{fig:workflow}) starts with network design. The \program{ANDL} is used to describe the required nodes, as well as the desired network topology, and the mapping of messages to different traffic classes. Afterwards our toolchain automatically generates an executable simulation configuration that is run using the simulation models for real-time Ethernet and fieldbusses. After the simulation run, the results are analysed with various result analysers that are built into the \program{\omnet} IDE, provided as additional plugins (e.g. the GCTA), or interconnected using databases and specialized output formats such as PCAPng.
\\
This section presents an example workflow of the simulation process -- from network description to result analysis.

\subsection{Network Modeling}
\label{subsec:network_modeling}
First step is the network description. This is done with the \program{ANDL} domain-specific language.
Listing \ref{lst:andl} shows an example of a network consisting of two CAN busses interconnected over a real-time Ethernet backbone described in the \program{ANDL}.

\begin{lstlisting}[language=ANDL,frame = lines,caption=\program{ANDL} code example with comments, label=lst:andl]
types std {            //Types can be defined and reused
  ethernetLink ETH {   //Definition for Ethernet link
    bandwidth 100Mb/s; //Link has bandwidth of 100MBit/s
  }
} //it is also possible to define types in a separate file

network smallNetwork { //network name is smallNetwork
  inline ini {         //Inline ini for special parameters
```
record-eventlog = false
```
  }                    //Parameters are inserted into .ini

  devices {            //Define all devices in the network
    ethernetLink eth1 extends std.ETH; //First Ethernet cable
    canLink cb1;       //First CAN bus
    canLink cb2;       //Second CAN bus
    node cn1;          //First CAN node
    node cn2;          //Second CAN node
    node en1;          //First Ethernet node
    node en2;          //Second Ethernet node
    gateway gw1 {      //Gateway for first CAN bus
    	pool gw1_1;      //Pool for Aggregation of CAN frames
    }
    gateway gw2;       //Gateway for second CAN bus
    switch s1;         //Real-time Ethernet Switch
  }

  connections {        //Physical connections (Segments = groups)
    segment backbone { //Ethernet Backbone part
      en1 <--> eth1 <--> s1;          //Ethernet Link
      en2 <--> {new std.ETH} <--> s1; //Ethernet Link
      gw1 <--> {new std.ETH} <--> s1; //Ethernet Link
      gw2 <--> {new std.ETH} <--> s1; //Ethernet Link
    }
    segment canbus {   //CAN bus part (busses share config)
      cn1 <--> cb1;    //CAN node connected to first bus
      gw1 <--> cb1;    //Gateway connected to first bus
      cn2 <--> cb2;    //CAN node connected to second bus
      gw2 <--> cb2;    //Gateway connected to second bus
    }
  }

  communication {      //Communication in the network
    message msg1 {     //First message definition
      sender cn1;      //First CAN node is sender
      receivers cn2;   //Second CAN node is receiver
      payload 6B;      //Message payload is 6 Bytes
      period 1ms;      //1ms cyclic transmission
      mapping {        //mapping to traffic class, id, gw strategy
        canbus: can{id 37;}; //Message ID 37 on CAN
        gw1: pool gw1_1{holdUp 2ms;}; //Aggregation time
        gw2;           //gw2 also responsible for the msg path
        backbone: tt{ctID 102;}; //TT traffic on backbone
  	  }
  	}
    message msg2 {     //Second message definition
      sender en1;      //First Ethernet node is sender
      receivers en2;   //Second Ethernet node is receiver
      payload 500B;    //Message payload is 500 Bytes
      period 125us;    //125us cyclic transmission
      mapping {        //mapping to traffic class
        backbone: avb{id 1;}; //AVB traffic on backbone
  		}
  	}
  }
}
\end{lstlisting}

The definition of the scenario starts with the required \code{devices} of the network.
The \code{connections} section arranges previously defined devices into a network topology.
This section shows also an additional way to instantiate \code{ethernetLink}.
It can be created in the specific link definition without the need of defining a name.
\\
The topology can be divided in several segments with different configurations for messages. In the example there is one segment for the Ethernet part called \code{backbone} and one segment for the CAN bus part called \code{canbus}.
A message traversing the border of a segment will be translated from the representation of the sending segment into the representation of the receiving one.
\\
The last part of the definition is the actual communication taking place. In the example there is one message transmitted from \code{cn1} to \code{cn2} and one message transmitted from \code{en1} to \code{en2}. The mapping of each message defines how the message is represented in the different segments. In the example the message \code{msg1} is a CAN frame with id \code{37} on the bus and a time-triggered message with critical traffic id \code{102} on the real-time Ethernet backbone.

Beside the features shown in this example, \program{ANDL} provides more parameters to describe traffic flows or aggregation strategies. Commonly used components can be defined in include files, e.g. a Ethernet Link with \SI{100}{\mega\bit\per\second}, and used in several places. Further \program{ANDL} provides inheritance, thus it is possible to define primitive stencils for components that are later refined during the instantiation.
\\
Currently, \program{ANDL} supports only the most common used simulation parameters. For more sophisticated configurations \code{inline ini} code can be used. Parameters defined in the \code{inline ini} sections are directly copied into the resulting \filename{.ini} files in an additional \code{with\_inline\_ini} configuration. Hence, \code{inline ini} definitions override generated definitions, which are placed in the \code{General} configuration.

\subsection{Experimentation}
\label{subsec:experimentation}
\begin{figure}[!ht]
\centering
\includegraphics[width=0.8\linewidth]{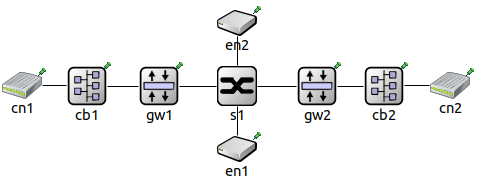}
\caption{\program{ANDL} generated network consisting of two CAN busses and a real-time Ethernet backbone with two gateways, two Ethernet nodes and one switch}
\label{fig:smallNetwork}
\end{figure}

In comparison to the compact description in \program{ANDL}, the size of the generated \omnet simulation configuration (.ini/.ned/.xml) has more than 250 lines. Nevertheless all relevant parameters for the in-car network designer are available in \program{ANDL}. The resulting network is shown in Figure \ref{fig:smallNetwork}.
\\
It is shown that the generation process has created the defined topology. Stimuli, TDMA scheduling and gateway strategies are generated, too. A simulation can be run immediately.
\\
Furthermore all \program{\omnet} features like distribution functions and configurations for concurrent simulations runs can be done in the \code{inline ini} part of the \program{ANDL} description.

Using \program{ANDL} in this way supports a fast experimentation process. Changes on topology, stimuli or gateway strategies are done in the \program{ANDL} and simulated in \program{\omnet} for an interactive analysis and comparison of different in-car network settings.

\subsection{Result Analysis}
\label{subsec:result_analysis}
\begin{figure}[!ht]
  \centering
  \includegraphics[width=1\textwidth]{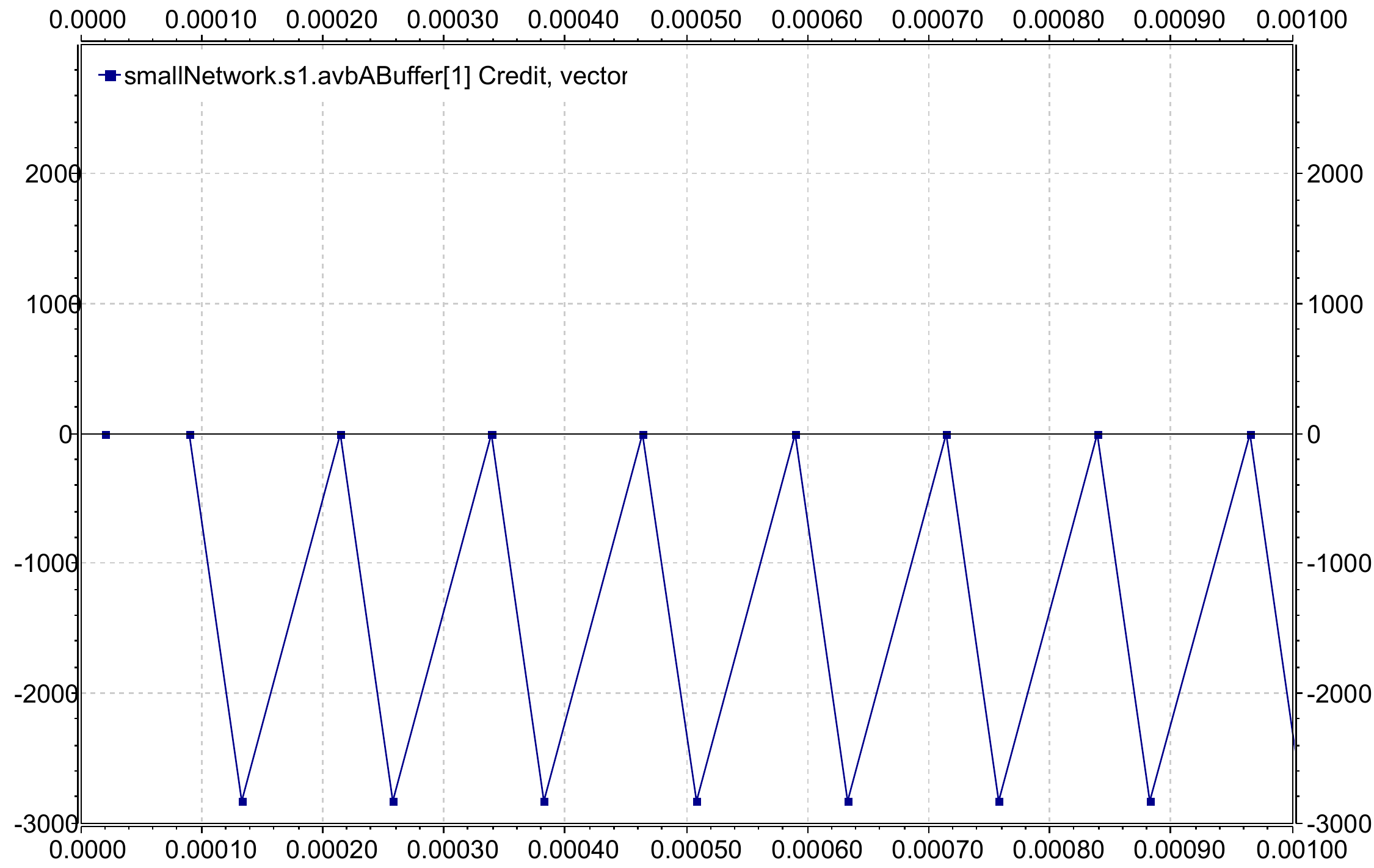}
  \caption{AVB Credit vector (\emph{s1} port 1) as seen in \program{\omnet}}
  \label{fig:omnetpp_credit}
\end{figure}

All known \program{\omnet} tools to analyse and visualize result data can be used to view in-car protocol specific results, too. To find the corresponding data for a devices defined in an \program{ANDL} description a user simply use the defined name to filter the result set. All user defined \program{ANDL} names are adopted to the \program{\omnet} simulation configuration.
An example for a protocol specific result is the credit of the AVB credit based shaper. Figure \ref{fig:omnetpp_credit} shows an extract of the credit vector for port 1 in switch \emph{s1}.

The \program{oppResultManagers} models enable distributed analysis of simulation results. It is realised using database ResultManagers. Listing \ref{lst:enableDatabaseResMan} shows how to enable database ResultManagers in a \filename{*.ini} configuration file of the simulation.

\begin{lstlisting}[language=OMNETPPINI,breaklines=true,frame=none,caption=Enabling database ResultManagers,label=lst:enableDatabaseResMan]
outputscalarmanager-class="cPostgreSQLOutputScalarManager"
outputvectormanager-class="cPostgreSQLOutputVectorManager"
postgresqloutputmanager-connection="dbname=testdb user=testuser password=testuser port=15432"
\end{lstlisting}

For example the \program{postgreSQL} manager allows to write simulation results on a central database server in the network, while simulations are executed on a distributed cluster of nodes. Several users can access the results concurrently without the necessity to distribute the result files. This allows to transfer the load of the simulation as well as result analysis from the users workstations towards strong servers and large centralized storage systems. The drawback of this solution is a slight performance decrease due to the overhead of sending results over the network as well as delays due to the databases lock mechanisms when it is accessed concurrently. Using a database system, \program{\omnet} simulations can be easily attached to a wide range of analysis tools, e.g. R \cite{rfgd-sdaer-08} using a database driver.

Using this way a ResultManager can be selected to fit the specific result analysis requirements.

%% file: case_study.tex
\section{Case Study: Automotive Backbone for Premium Cars}
\label{sec:case_study}
This section presents a system level simulation case study based on an anonymized communication matrix of the Volkswagen Golf 7 that support the MQB platform of Volkswagen.
It is extended by high bandwidth communications that transport measurements of two cameras and two lidars to support features like sensor fusion based on raw data.
This case study demonstrates the relevance of system level simulation within the field of future in-car network designs, gives an impression of how to use and handle the simulation tools, and finally provides selected results obtained by this case study.
All results are the outcome of \program{\omnet} simulations.
The case study was supported by the German Federal Ministry of Education and Research (BMBF) under the project RECBAR \cite{smkr-reicb-14}.

\subsection{Case Study \& Metrics}
\begin{figure}[!ht]
\centering
\includegraphics[width=\linewidth]{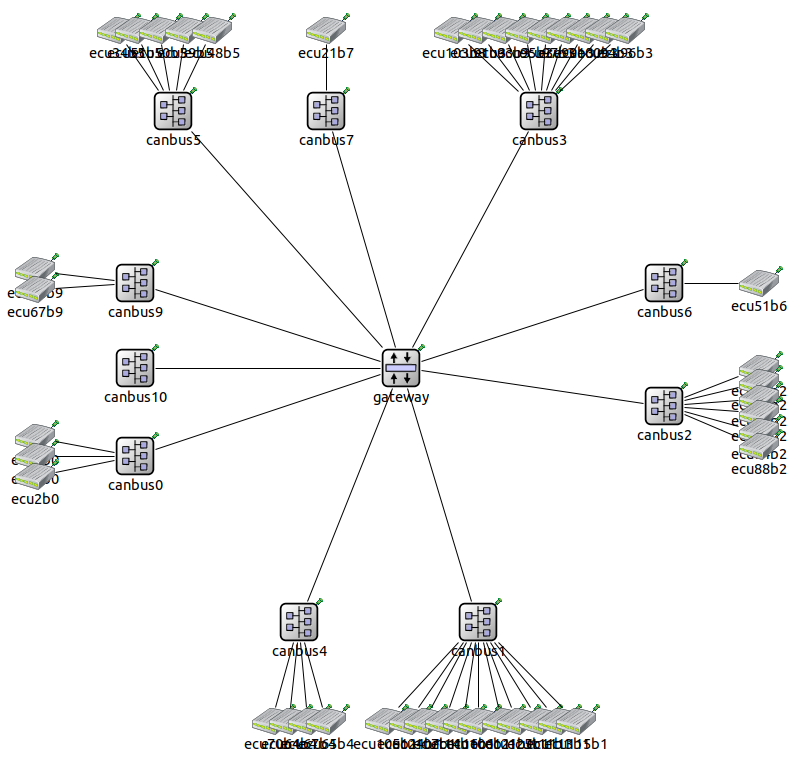}
\caption{Central CAN gateway design}
\label{fig:centralCANGateway}
\end{figure}

To get results for comparison the first simulation analyses the current CAN based communication structure of premium cars (s. \autoref{fig:centralCANGateway}).
It is based on an central CAN gateway that connects the domain specific CAN buses and realizes the exchange of CAN frames between these buses.
The design consists of seven public and two private CAN buses.
All ECUs connected to this buses and the corresponding periodic CAN traffic will be simulated (s. \autoref{tab:ecu_canbus} and \autoref{tab:perioden}).
CAN traffic based on acyclic messages will not be simulated.

Even if the CAN identifiers have been anonymized, the prioritization remains in accordance with the original MQB communication matrix.

\begin{figure}[!ht]
\includegraphics[width=\linewidth]{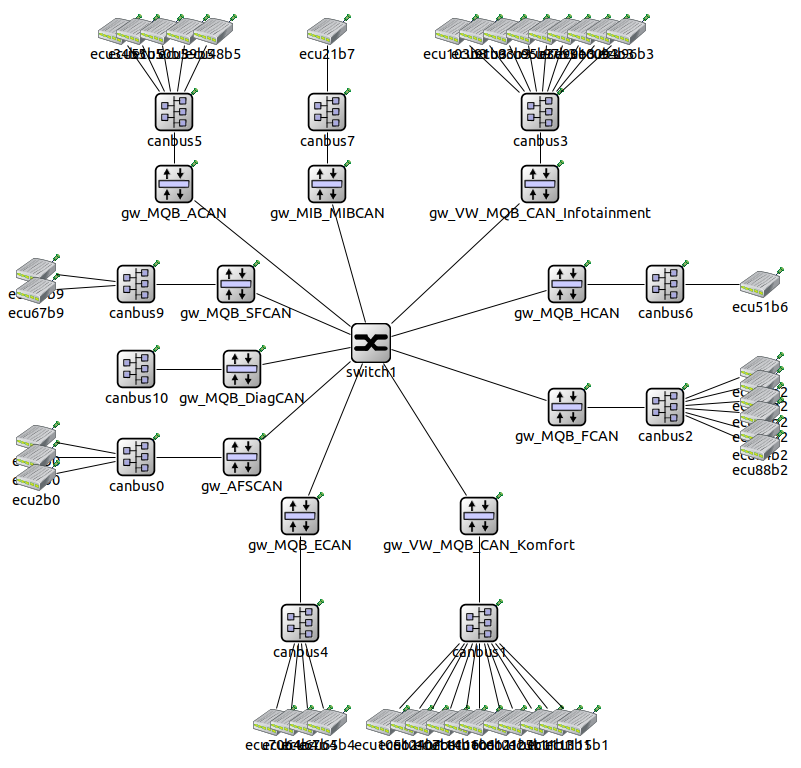}
\caption{One Ethernet switch design}
\label{fig:oneEthernetSwitchDesign}
\end{figure}

\begin{figure}[!ht]
\includegraphics[width=\linewidth]{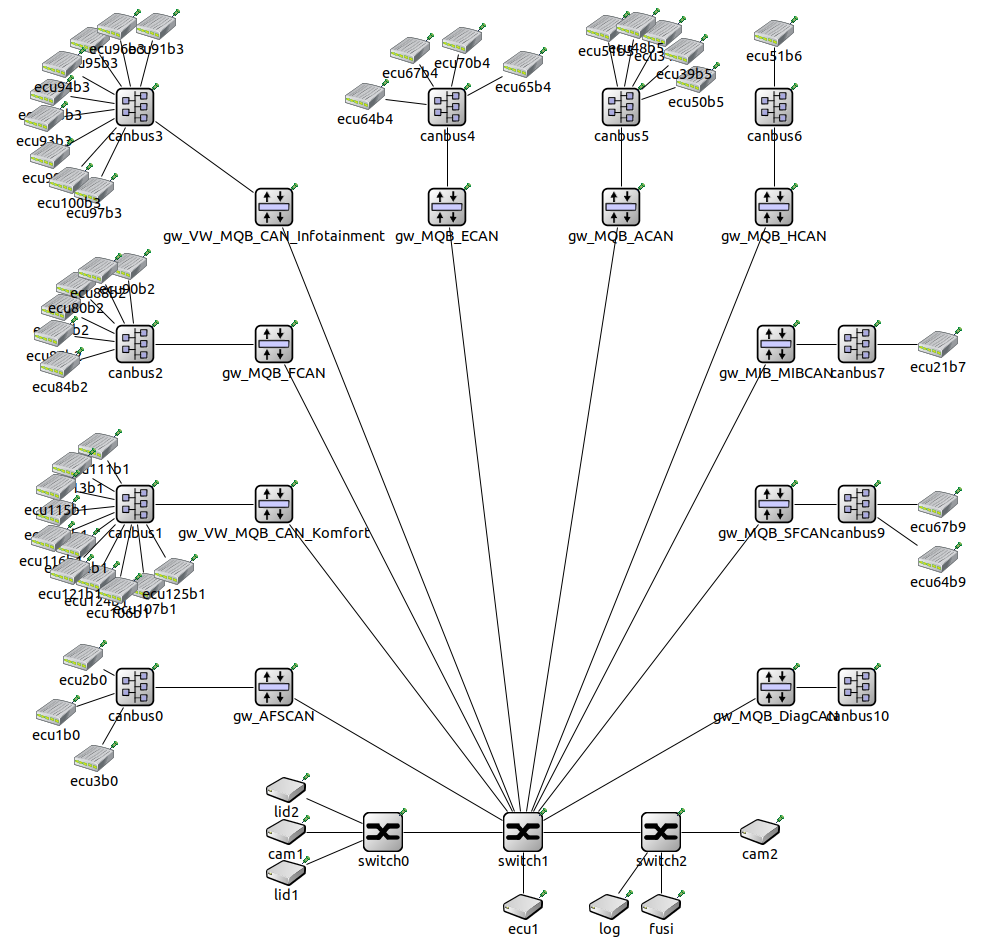}
\caption{Ethernet backbone within the RECBAR car}
\label{fig:recbarEthernetDesign}
\end{figure}

\begin{table}
\caption{Number of ECUs per CAN bus}
\label{tab:ecu_canbus}
\begin{tabular*}{0.48\linewidth}{@{\extracolsep{\fill}}l S[table-format=2.0, round-precision=0]}
	\toprule
	{CAN bus} 	& {Number of ECUs}\\
	\midrule
    canbus0        & 3\\
    canbus2        & 6\\
    canbus4        & 4\\
    canbus6        & 1\\
    canbus9        & 2\\
  	\bottomrule \addlinespace[0.5em]
\end{tabular*}
\hfill
\begin{tabular*}{0.48\linewidth}{@{\extracolsep{\fill}}l S[table-format=2.0, round-precision=0]}
	\toprule
	{CAN bus} 	& {Number of ECUs}\\
	\midrule
    canbus1   & 11 \\
    canbus3   & 9  \\
    canbus5   & 5  \\
    canbus7   & 1  \\
    canbus10  & 0  \\
  	\bottomrule \addlinespace[0.5em]
\end{tabular*}
\end{table}
\begin{table}
\caption{Number of CAN Messages per period}
\label{tab:perioden}
\begin{tabular*}{0.48\linewidth}{@{\extracolsep{\fill}} S[table-format=4.0, round-precision=0] S[table-format=2.0, round-precision=0]}
	\toprule
	{Period [\si{\ms}]} 	& {Number of messages}\\
	\midrule
  10 	  & 23\\
  25      &  1\\
  40      & 33\\
  60      &  2\\
  100     & 69\\
  160     &  1\\
  450     &  2\\
  1000    & 38\\
  	\bottomrule \addlinespace[0.5em]
\end{tabular*}
\hfill
\begin{tabular*}{0.48\linewidth}{@{\extracolsep{\fill}} S[table-format=4.0, round-precision=0] S[table-format=2.0, round-precision=0]}
	\toprule
	{Period [\si{\ms}]} 	& {Number of messages}\\
	\midrule
  20      & 38 \\
  30      &  1 \\
  50      & 17 \\
  80      & 10 \\
  150 	  & 10 \\
  200     & 50 \\
  500     & 49 \\
  2000    &  3 \\
  	\bottomrule \addlinespace[0.5em]
\end{tabular*}
\end{table}
In the second simulation, the central CAN gateway is replaced by an Ethernet switch and CAN to Ethernet gateways (s. \autoref{fig:oneEthernetSwitchDesign}).
The final simulation analyzes a network that consists of real-time Ethernet backbone using three real-time switches and several additional nodes with high bandwidth applications such as high definition cameras and laser scanners (s. \autoref{fig:recbarEthernetDesign}).
They represent first steps of the gradual transition to a flat Ethernet topology.

The simulation records the following metrics:

\begin{itemize}
\item {\bf Utilized bandwidth} This is the bandwidth used on all Ethernet links and CAN buses, including all protocol overheads.

Results can be obtained from the scalars \cursive{bitsPerSec} of the \cursive{canBusLogic} module and \cursive{bits/sec sent} or \cursive{bits/sec rcvd} of the \cursive{mac} module.
\item {\bf Latency} The simulation records end-to-end latency across the entire route between sending and receiving ECU.
It includes the time spent within gateways.
The time starts at the point in time when a data source provides a package for sending. It ends at the time of arrival of the package at a data sink.
If applicable, the latency on the Ethernet network is investigated - without CAN bus transmission. This highlights the effects of Ethernet configurations.

The latency will be recorded for each ID at each receiver and each station on the path from the sender to receiver.
Since a large amount of data is generated in the simulation, this section gives only maximum and minimum latencies.

Many modules in our simulation models have the \cursive{rxLatency} vector results. For End-to-End latency we are looking at the consuming applications.
\item {\bf Jitter} This metric is defined for periodic messages.
It is the absolute value of the maximum difference of the latency belonging to consecutive messages.
For messages sent from different points in the network to the same sink the jitter is ignored.
\item {\bf Queue Length} The simulation observes the length of all queues in the network.
Moreover the packet lost due to buffer overflow will be monitored. This is also important for the design of the target hardware.

For obtaining the queue length statistics we have to look for the \cursive{QueueLength} vectors.
\end{itemize}

\subsection{Central CAN gateway design}
This scenario depicts the series configuration of the vehicle and simulates the initial situation without Ethernet backbone.
This represents the actual communication topology of many current cars.
The recorded metrics serve as reference for further architectural variants and configurations.
\Autoref{fig:centralCANGateway} shows the simulation configured in \program{\omnet}.
Each CAN bus is connected to the central CAN Gateway.
Inter-domain traffic is routed through this gateway.
\\
There are a total of 416 different messages on the buses, of which 216 are transported via the gateway.
The central CAN gateway delays messages by \SI{60}{\us}.
This value corresponds to an average value measured in the real vehicle.

\begin{table}
\caption{Utilized bandwidth: analytical vs. simulation results}
\label{tab:compareCanBandwidth}       
\begin{tabular*}{\linewidth}{@{\extracolsep{\fill}} l S[table-format=3.3, round-precision=0] S[table-format=3.3, round-precision=0] S[table-format=1.3, round-precision=0]}
	\toprule
	{Bus} & {Analytical result [\si{\kilo\bit\per\second}]}  & {Simulation result [\si{\kilo\bit\per\second}]} & {Deviation [\percent]} \\
	\midrule
	canbus0 &  97,919  &  97,918 & 0,001 \\
	canbus1 & 128,130  & 127,329 & 0,625 \\
	canbus2 & 237,788  & 232,225 & 2,339 \\
	canbus3 &  63,125  &  63,121 & 0,006 \\
	canbus4 & 113,238  & 113,232 & 0,005 \\
	canbus5 & 221,631  & 220,509 & 0,506 \\
	canbus6 &   5,215  &   5,214 & 0,019 \\
	canbus7 &   2,801  &   2,800 & 0,036 \\
	canbus9 & 184,602  & 184,589 & 0,007 \\
  	\bottomrule \addlinespace[0.5em]
\end{tabular*}
\end{table}

\begin{table}
\caption{Exemplary end-to-end latencies}
\label{tab:latencyCentralCanGateway}
\begin{tabular*}{\linewidth}{@{\extracolsep{\fill}} l S[table-format=5.3, round-precision=0] S[table-format=4.3, round-precision=0]}
	\toprule
	{CAN-ID} & {Maximum end-to-ende latency [\si{\us}]} & {Average end-to-ende latency [\si{\us}]} \\
	\midrule
	 17  &   946,707 & 572,445 \\
	 331  &  8465,906 & 845,920 \\
	 510  & 17974,989 & 1168,952 \\
  	\bottomrule \addlinespace[0.5em]
\end{tabular*}
\end{table}

For simulation model validation the utilized bandwidth was determined analytically based on the cycle times given in the communication matrix.
\Autoref{tab:compareCanBandwidth} compares analytical and simulation results.
Due to a different recording technique the deviation is up to \SI{2,339}{\percent}.
The analytical approach is based on the cycle times, solely.
In contrast, the simulation takes only packages into account that are transmitted over the bus.
Packages that are still waiting for transmission in the gateway queue or packages that are waiting for the access to the bus are not included here.
Therefore simulated bandwidth lies slightly below the analytically calculated.

End-to-end latencies for all CAN messages were determined.
\Autoref{tab:latencyCentralCanGateway} gives some exemplary results.
They show the great influence of CAN message priority on latency.
This is because of the CAN bus arbitration which always favors the message with the lowest CAN ID value.
While the maximum latency of the highest prioritized message (lowest CAN ID value) is less than \SI{1}{\ms}, it rises to almost \SI{18}{\ms} for the lowest prioritized message (highest CAN ID value).
This effect affects the average latency, too.

The same effect occurs for the jitter of CAN messages.
Typically, the jitter of low prioritized CAN messages is higher.
For the analysis of the jitter, a distinction must be made between messages that are transmitted locally on a CAN bus and messages that are transmitted via the central gateway.
For each CAN bus \autoref{tab:jitterCentralCanGateway} gives the jitter for CAN messages consumed on this bus.
As a result, messages that are forwarded via the central gateway do not necessarily have a higher jitter.
Rather, the jitter is a mix of effects in the gateway, arbitration on multiple buses, and message prioritization.
For example on canbus9 the minimum jitter of local messages is higher than messages that traverse the gateway due to a higher prioritized CAN message.

\begin{table}
\caption{Comparison of minimal and maximal jitter}
\label{tab:jitterCentralCanGateway}
\begin{tabular*}{\linewidth}{@{\extracolsep{\fill}} l S[table-format=1.3, round-precision=0] S[table-format=2.3, round-precision=0]  S[table-format=1.3, round-precision=0] S[table-format=2.3, round-precision=0]}
	\toprule
	        & \multicolumn{2}{c}{local}     & \multicolumn{2}{c}{via gateway} \\
	        \cmidrule(lr){2-3} \cmidrule(lr){4-5}
	{Bus}   & {Min [\si{\ms}]}& {Max [\si{\ms}]} & {Min [\si{\ms}]} & {Max [\si{\ms}]} \\
	\midrule
	canbus0 & 0,574         &  7,511               & 4,301         &  5,866 \\
	canbus1 & 1,910         & 30,629               & 0,983         & 16,295 \\
	canbus2 & 1,504         & 15,827               & 0,961         & 23,196 \\
	canbus3 & 1,702         &  9,430               & 0,935         & 23,995 \\
	canbus4 & 2,154         &  9,735               & 0,957         & 24,505 \\
	canbus5 & 0,564         & 19,383               & 1,868         & 20,044 \\
	canbus6 & \multicolumn{2}{c}{no local traffic} & 8,017         & 19,383 \\
	canbus7 & \multicolumn{2}{c}{no local traffic} & 5,278         & 15,920 \\
	canbus9 & 4,346         & 25,641               & 0,959         & 23,707 \\
  	\bottomrule \addlinespace[0.5em]
\end{tabular*}
\end{table}

For an architecture with a central CAN gateway these measurements show that the arbitration of the CAN messages has a significantly higher impact on jitter and latency than the delay caused by the gateway (\SI{60}{\micro\second}).
This is a good starting position for architectures based on an Ethernet backbone.
\subsection{One Ethernet Switch design}
This architecture is an intermediate step to an architecture with Ethernet backbone.
The central CAN gateway is replaced by an Ethernet switch and CAN - Ethernet gateways (s. \autoref{fig:oneEthernetSwitchDesign}).
This simulation uses the same stimuli (CAN messages) as the previous simulation.

Each CAN bus is connected to its own CAN-Ethernet gateway.
In accordance with the behavior of the CAN central gateway, each CAN message, which flows via the central CAN gateway in the original design, flows via the Ethernet switch in this architecture.
On the Ethernet side it is sent from the source gateway to the destination gateway.
A CAN message maybe forwarded to several CAN buses.
In this case the simulation implements this behavior by sending a separat frame for each destination.
Alternatively, it could be realized using multicast.

The Ethernet frames will be sent via the standard Ethernet protocol (best effort) using \SI{100}{\mega\bit\per\second} links.
In the current simulation, the processing delay of a gateway is \SI{40}{\us}.
This value is based on measurements of an ARM-9 based prototype gateway.

The maximum payload of a CAN frame is \SI{8}{\byte}. The minimal payload of an Ethernet frame is \SI{46}{\byte}.
Without aggregation of multiple CAN messages within one Ethernet frame, Ethernet bandwidth would be wasted.

\subsubsection*{One Ethernet Switch Design without Aggregation}
In general, the utilized bandwidth on CAN buses does not change compared to the central Ethernet gateway design.
The bandwidth used on the Ethernet links is always below \SI{1}{\mega\bit\per\second}, but above that utilized bandwidth on CAN buses.
This is due to the lack of aggregation of CAN messages.
By padding and the larger size of the Ethernet frames compared to CAN frames, the necessary bandwidth increases on the Ethernet.
It should be noted that no multicast is used and therefore messages that are transmitted to multiple buses are also sent multiple times.

The maximum end-to-end latency is \SI{25,289}{ms}.
The minimum end-to-end is \SI{0,126}{ms} (local on a CAN bus).
The delay on the Ethernet network is always less than \SI{10}{\percent} of the total end-to-end latency.
The latency on the Ethernet from transmitter to receiver via a switch (\SI{8}{\micro\second} hardware delay) is between \SI{19,92}{\us} and \SI{60,24}{\us}.

The queues in front of the output links of the switch store a maximum of 2 frames.
The queues of the gateways stores frames that are waiting to be processed by the CAN side of the gateway.
These queues store a maximum of 6 Ethernet frames.

In summary, this simulation shows that the use of an Ethernet switch has little effect on the transmission of CAN messages.
The latency and jitter metrics continue to be significantly influenced by CAN bus arbitration.

\subsubsection*{One Ethernet Switch Design with Aggregation}
This section investigates the aggregation of CAN messages into one Ethernet frame.
It is based on the approach that incoming CAN messages will be buffered in the gateway, so that ultimately several CAN messages will transmitted in one Ethernet frame to reduce the utilized bandwidth on the Ethernet.
On the other hand the delay in a gateway increase CAN message latencies.
\begin{figure}[!ht]
\includegraphics[width=\linewidth]{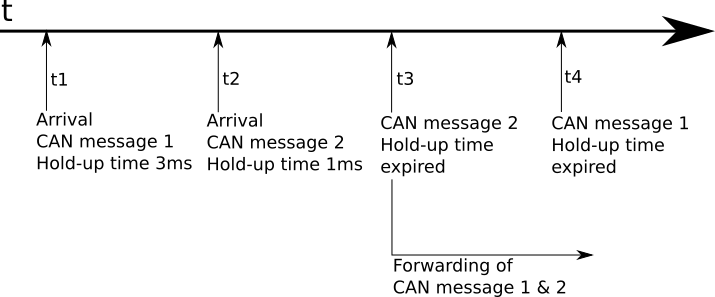}
\caption{Aggregation of CAN messages with an pool}
\label{fig:CANMessageAggregation}
\end{figure}

The gateway implements aggregation of CAN messages as follows:
To each CAN message that arrives at a gateway a hold-up time is assigned.
It defines how long a message can be buffered in the gateway until it is sent to the Ethernet.

CAN messages will be aggregated in buffers called pools. A pool stores all messages that will be aggregated in the same Ethernet frame.
As soon as the hold-up time of one CAN message expires, all buffered CAN messages of the corresponding pool will be passed to the transformation module of the gateway.
For each target gateway the transformation module sends an Ethernet frame that contains the CAN messages stored in the pool.
\Autoref{fig:CANMessageAggregation} shows how such a pool works. At time $t_3$, the hold-up time of CAN message 2 expires.
Therefore, all messages that have arrived so far will be forwarded at this time.

There are two ways to configure pool aggregation.
One possibility is to use the \program{ANDL} like in \autoref{lst:andl} in \autoref{subsec:network_modeling}.
Another possible way is to manipulate the generated \filename{.xml} file.
Examples of this can be seen in the examples folder of the \program{SignalsAndGateways} model.
\begin{table}
\caption{Initial pool configuration}
\label{tab:holdUpTimesBasedOnCANId}
\begin{tabular}{p{2cm}p{5cm}}
	\toprule
	CAN-ID 		& Hold-up time scenario 1 \\
	< 101     & \si{0}{ms} \\
	101 - 200 & \si{25}{\%} of the period of CAN-ID \\
	201 - 300 & \si{50}{\%} of the period of CAN-ID \\
	300 < 	  & \si{75}{\%} of the period of CAN-ID \\
	\bottomrule \addlinespace[0.5em]
\end{tabular}
\end{table}
\\
Three different aggregation scenarios will be simulated.
\begin{enumerate}
\item Within this scenario hold-up times are based on the CAN-ID and period of the CAN message.
Each hold-up time is calculated from the period and the percentage given in \autoref{tab:holdUpTimesBasedOnCANId}.
All CAN messages are stored in the same pool.
Hence, high-priority CAN messages are delayed less than low-priority messages.
\item This scenario differs from the first one in the aspect that messages with high priority CAN messages (CAN-IDs <101) get a hold-up time of \SI{1}{\ms}.
This increases the likelihood that high priority CAN message will be aggregated with other CAN messages.
\item This scenario uses 2 pools. CAN messages with similar hold-up times belong to the same pool.
\end{enumerate}

\begin{figure}[!ht]
\includegraphics[width=\linewidth]{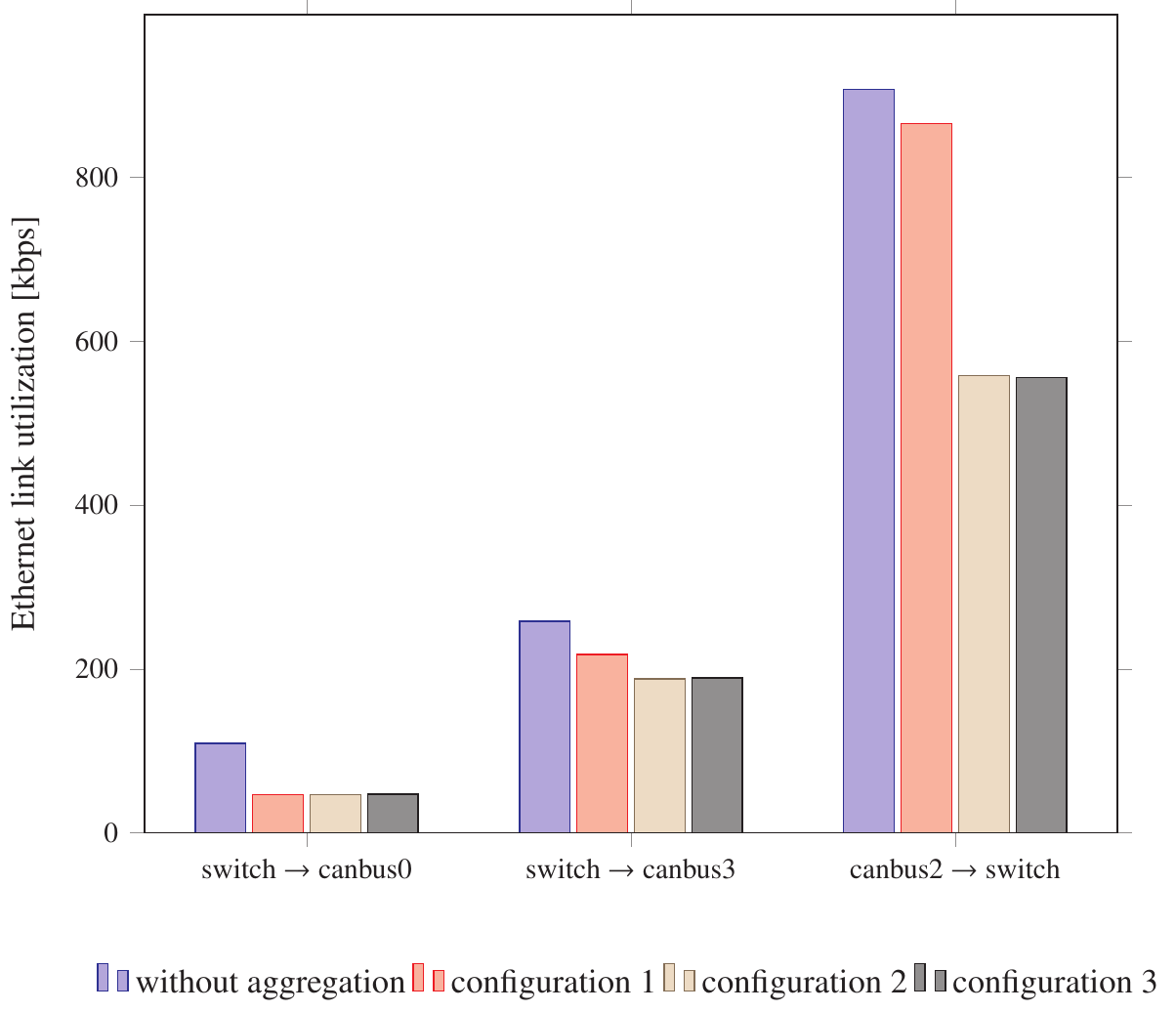}
\caption{Utilized bandwidth on 3 Ethernet links}
\label{fig:utilized_bandwidth}
\end{figure}

\Autoref{fig:utilized_bandwidth} represents the utilized bandwidth for three Ethernet links.
As expected, it shows that aggregation reduces utilized Ethernet bandwidth.
Depending on the structure of the traffic on the buses, the utilized bandwidth is reduced by more than \SI{50}{\percent}.
The difference between the three configurations is small.
It is worth noting that none of the configurations provides the lowest utilized bandwidth for all links.
In contrast to the Ethernet links aggregation does not change the utilized bandwidth on CAN buses.

\begin{table}
\caption{Maximum End-to-End latency for some CAN-IDs on canbus1}
\label{tab:latencyWithAggregation}
\begin{tabular}{p{1cm}p{1,8cm}p{1,8cm}p{1,8cm}p{1,8cm}p{1,8cm}}
	\toprule
     CAN-ID & central     & without     & \multicolumn{3}{c}{with aggregation}\\
            & CAN gateway & aggregation & configuration 1 & configuration 2 & configuration 3 \\
            & [\si{us}] & [\si{us}] & [\si{us}] & [\si{us}] & [\si{us}] \\
	\midrule
     17 &   946,707 &   984,859 &   990,248 &  1984,378 &   1987,647 \\
		 331 &  8465,906 &  8658,725 & 12835,177 & 13643,712 &  16676,314 \\
		 510 & 17974,989 & 18415,130 & 23217,447 & 24470,504 & 754554,309 \\
	\bottomrule \addlinespace[0.5em]
\end{tabular}
\end{table}

\begin{table}
\caption{Number of CAN Messages within a pool}
\label{tab:messagesAggregationInAnEthernetFrame}
\begin{tabular}{p{1,4cm}p{1,4cm}p{1,4cm}p{1,4cm}p{1,4cm}p{1,4cm}p{1,4cm}}
	\toprule
     CAN bus  & \multicolumn{2}{c}{configuration 1} & \multicolumn{2}{c}{configuration 2} & \multicolumn{2}{c}{configuration 3} \\
              & max       & average       & max       & average       & max & average \\
     \midrule
     canbus0 &  5 & 1,44 &  5 & 1,44 &  3 & 1,41 \\

		 canbus1 & 23 & 5,03 & 23 & 5,20 & 12 & 4,59 \\
     canbus2 & 14 & 1,23 & 16 & 2,81 & 16 & 2,80 \\

		 canbus3 & 13 & 9,09 & 13 & 9,08 &  9 & 6,81 \\

		 canbus4 & 17 & 2,49 & 19 & 5,15 & 19 & 5,16 \\

		 canbus5 & 18 & 1,19 & 21 & 2,70 & 17 & 2,69 \\

		 canbus6 & \multicolumn{6}{c}{gateway does not send} \\

		 canbus7 & \multicolumn{6}{c}{gateway does not send} \\

		 canbus9 & \multicolumn{6}{c}{gateway does not send} \\
  	\bottomrule \addlinespace[0.5em]
\end{tabular}
\end{table}

As expected, aggregation increases the end-to-end latency.
\Autoref{tab:latencyWithAggregation} gives some latencies of aggregated messages.
Configuration 1 provides best performance.
For CAN ID 17 latency is similar to the one without aggregation.
This is because of the hold-up time value of \SI{0}{\milli\second} for CAN IDs < 101.
Within configuration 2 this latency increases by the configured hold-up time of \SI{1}{\ms}.
In the third configuration, the latency of CAN-ID 510 increases significantly.
This is due to the subdivision into two different pools depending on hold-up time.
The probability that the pool forwarding is triggered, by a CAN message with a faster expired hold-up time, is much lower.
This can be verified by inspection of the \cursive{HoldUpTime} vector in the gateway buffering module.

Regarding to jitter, CAN messages that are not transported via a gateway have little difference between the configuration with and without aggregation.
This difference is based on bursts generated by gateways as follows:
If an Ethernet frame containing several CAN messages arrives at a gateway, it fills the message object buffers of CAN bus interface with CAN frames of different priority at the same time.

\begin{figure}[!ht]
\includegraphics[width=\linewidth]{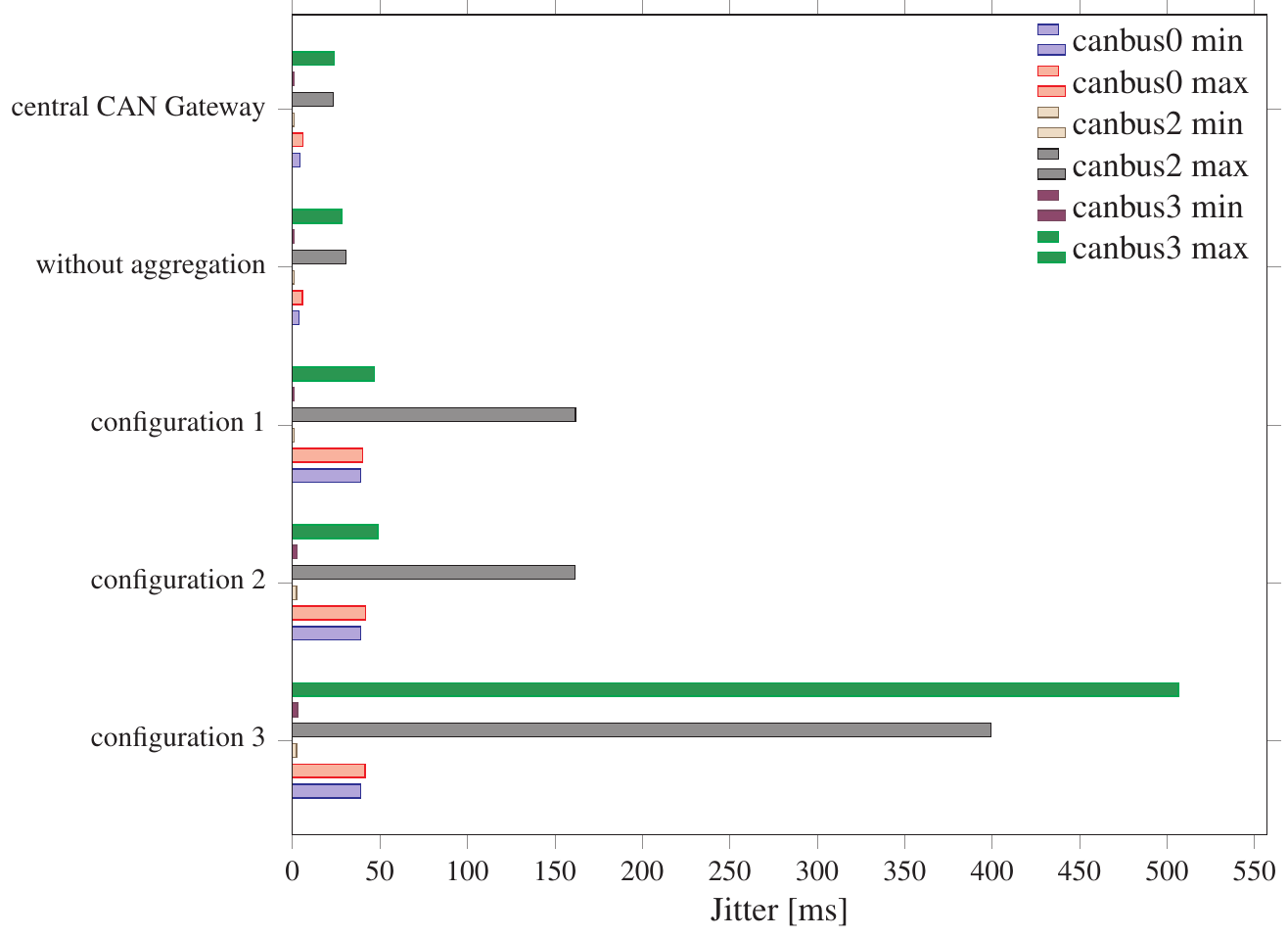}
\caption{Minimal and maximal jitter on 3 Ethernet links}
\label{fig:jitter_plot}
\end{figure}

\Autoref{fig:jitter_plot} gives the minimal and maximal jitter for messages transported via  gateways for three can buses.
Aggregation leads to an significant higher maximum jitter compared to the configuration without aggregation.
In particular, this concerns low prioritized ID messages, due to the bursts on the destination bus and their long hold-up time.

In terms of aggregation performance, the number of messages that are aggregated in an Ethernet frame is of special interest.
\Autoref{tab:messagesAggregationInAnEthernetFrame} represents this metric.
With up to 23 aggregated CAN messages, some of the pools are very large.
However, far fewer CAN messages will be aggregated on average.
Thus, the aggregation configurations provide potential for further optimization.
A general statement about an optimal pool strategy can not be given.

In summary, the results of this section show a trade-off between bandwidth on the one side and latency / jitter on the other side.
The respective aggregation strategy has a massive impact on latency, jitter and bandwidth.
For some aggregation strategies, the gain in bandwidth is particularly efficient relative to the effects on latency and jitter.

\subsection{Realtime Ethernet Backbone design}
In this scenario the Ethernet backbone communication architecture of the real RECBAR prototype car \cite{smkr-reicb-14} is simulated.
In addition to the previous communication matrix, additional network participants has been added.
This components are a front- and a rear-camera, two lidar in the front of the car, a logging ECU and an ECU for sensor fusion based on raw data.
ECU1 periodically sends synchronization frames for the global time synchronization.
The backbone contains three Ethernet switches.
All links have a bandwidth of \SI{100}{\mega\bit\per\second}.

To add this additional devices to the network configuration we use the flexibility gained by the \program{ANDL}. To extend the topology just the lines shown in \autoref{lst:andl_recbar} are added to the existing \program{ANDL} model.

\begin{lstlisting}[language=ANDL,frame = lines,caption=\program{ANDL} code example with comments, label=lst:andl_recbar]
...
connections {
    segment backbone {
        lid1 <--> {new std.ETH100M} <--> switch0;
        lid2 <--> {new std.ETH100M} <--> switch0;
        cam1 <--> {new std.ETH100M} <--> switch0;
        switch0 <--> {new std.ETH100M} <--> switch1;
        ecu1 <--> {new std.ETH100M} <--> switch1;
        gateway0 <--> {new std.ETH100M} <--> switch1;
        gateway1 <--> {new std.ETH100M} <--> switch1;
        gateway2 <--> {new std.ETH100M} <--> switch1;
        gateway3 <--> {new std.ETH100M} <--> switch1;
        gateway4 <--> {new std.ETH100M} <--> switch1;
        gateway5 <--> {new std.ETH100M} <--> switch1;
        gateway6 <--> {new std.ETH100M} <--> switch1;
        gateway7 <--> {new std.ETH100M} <--> switch1;
        gateway9 <--> {new std.ETH100M} <--> switch1;
        gateway10 <--> {new std.ETH100M} <--> switch1;
        switch1 <--> {new std.ETH100M} <--> switch2;
        cam2 <--> {new std.ETH100M} <--> switch2;
        log <--> {new std.ETH100M} <--> switch2;
        fusi <--> {new std.ETH100M} <--> switch2;
    }
    segment canbus {
        ecu3b0 <--> canbus0;
        ...
    }
}
...
\end{lstlisting}

In addition new message definition must be added or extended with new receivers. The generation of this extended \program{ANDL} results into the new target topology and message definitions.

The CAN gateways are connected to the central switch.
In addition to the previous CAN communication, the CAN messages are sent to a logging ECU that is connected to \code{switch2}.
Due to low camera resolution and compressed data transmission each camera stream requires \SI{7}{\mega\bit\per\second} bandwidth.
We use 2D four layer laser scanners (lidar). Hence a lidar requires \SI{2,5}{\mega\bit\per\second} bandwidth.
These streams are sent to the logging unit and a fusion calculator.

The communication bases on rate-constrained traffic class (AFDX).
This traffic class uses multicast.
The biggest link load is between \code{switch2} and the logging unit (about \SI{20,2}{\mega\bit\per\second}).

\begin{table}
\caption{End-to-End latency for some CAN-IDs on canbus1 of the RECBAR car}
\label{tab:latencyRecbarCar}
\begin{tabular}{p{1,2cm}p{3cm}p{3cm}}
	\toprule
     CAN-ID & max. latency [\si{\us}] & average latency [\si{\us}] \\
	\midrule
      17 &  532,264 & 331,641 \\
			331 & 4191,718 & 380,729 \\
			510 & 8625,544 & 595,964 \\
	\bottomrule \addlinespace[0.5em]
\end{tabular}
\end{table}

In contrast to previous simulations, rate constrained traffic will be used.
Therefore the utilized bandwidth decreases due to multicast forwarding.
Again this can be investigated through the \cursive{QueueSize} vectors. When a frame is forwarded to multiple destinations just on frames is queued instead of one per target.

Comparing the numbers of \autoref{tab:latencyWithAggregation} and \autoref{tab:latencyRecbarCar} shows hat this configurations reduces the end-to-end latency of CAN messages.
This is due to the use of multicast.
At the sending gateway no more frames, which different unicast destination addresses and the same payload, appear at the same time.

In summary, the simulation results of this scenario show that latency and jitter are more affected by CAN bus arbitration than by the Ethernet backbone architecture.
It can be seen that multicast addressing saves the bandwidth of corresponding Ethernet links and reduces the latency, too.

%% file: conclusion.tex
\section{Conclusion and Outlook}
\label{sec:conclusion}

Automotive industry is currently re-thinking the communication technologies in cars, thereby developing a strong preference towards real-time Ethernet. The design and evaluation of future Ethernet-centric architectures, but more delicately the transition  from  current legacy busses and gateways to a distributed switching layer requires extensive experimentation and careful evaluation of every design step. This work largely profits from available simulation tools and platforms that allow for a rapid assessment of design choices with high accuracy.

We presented an environment for modeling and simulating of future in-car networks based on the \program{\omnet} simulator and its \program{INET} framework. This suite includes  simulation models for various real-time extensions of Ethernet including AVB and TSN, field-busses including CAN and FlexRay, Gateways, as well as tools for modeling vehicular networks. This rich environment enables researchers from academia and developers from industry to thoroughly investigate network concepts and designs that are composed of the current and the emerging link-layer technologies.

As a special  support for engineers during their design process, the domain specific language \program{ANDL} has been defined. It allows to describe design variant on an abstract level and supports a fast exploration of different design variants. We illustrated its utility with  a case study based on realistic automotive data and demonstrated the practicability of this approach.

In future work, we will proceed  in three directions. (1) Within the automotive industry, detailed support of layer 3 and 4 protocols (like IEEE 1722, TCP and UDP) on top of the QoS-enhanced link layer is of rising interest. Therefor a interface has to be implemented into our models, so that existing models of higher layers are easily adaptable. (2) The TSN design team investigates pre-emption, and we plan on integrating different variants and investigating them  in our models. (3) Current applications allow stimuli generation based on random numbers. Domain specific reactive behavior could make the stimuli generation even more realistic. Corresponding concepts, adapted to real-time protocols will be introduced.

%% file: acronyms.tex
%
\section*{List of Acronyms}
\label{sec:acronyms}
\acresetall		
\begin{multicols}{2}	
\begin{acronym}[6LoWPAN]
	\acro{6LoWPAN}{\acl{IPv6} over Low power Wireless Personal Area Network}%
	\acro{6TiSCH}{\acl{IPv6} over the \acl{TSCH} mode of \acl{IEEE} 802.15.4e}%
	\acro{ACC}{Adaptive Cruise Control}%
	\acro{ACK}{Acknowledgment}%
	\acro{ADAS}{Advanced Driver Assistance System}\acroplural{ADAS}[ADAS]{Advanced Driver Assistance Systems}%
	\acro{ACT}{Allocation Counter Table}%
	\acro{ADC}{Analog-to-Digital Converter}%
	\acro{AE}{Application Entity}%
    \acro{AEI-id}{Application Entity Instance \acl{ID}}%
	\acro{AEN}{Application Entity Name}%
	\acro{AES}{Advanced Encryption Standard}%
	\acro{AH}{Authentication Header}%
	\acro{AIFS}{Arbitration Inter-Frame Spacing}%
	\acro{AMC}{Adaptive Modulation and Coding}%
	\acro{AMQP}{Advanced Message Queuing Protocol}%
	\acro{ANDL}{Abstract Network Description Language}%
	\acro{ANSI}{American National Standards Institute}%
	\acro{AODV}{Ad-hoc On-Demand Distance Vector}%
	\acro{AP}{Application Process}%
    \acro{API}{Application Programming Interface}%
	\acro{API-id}{Application Process Instance \acl{ID}}%
	\acro{APN}{Application Process Name}%
	\acro{ARP}{Address Resolution Protocol}%
	\acro{ASCII}{American Standard Code for Information Interchange}%
	\acro{ASE}{Attack Simulation Engine}%
	\acro{ASH}{Auxiliary Security Header}%
	\acro{ASI}{Attack Specification Interpreter}%
	\acro{ASK}{Amplitude Shift Keying}%
	\acro{ASL}{Attack Specification Language}%
	\acro{ASN.1}{Abstract Syntax Notation One}%
	\acro{AVB}{Audio Video Bridging}%
	\acro{BAN}{Body Area Network}%
	\acro{BATMAN}{Better Approach To Mobile Adhoc Networking}%
	\acro{BDD}{Behavior Driven Development}%
	\acro{BE}{Backoff Exponent}%
	\acro{BER}{Bit-Error Rate}%
	\acro{BGP}{Border Gateway Protocol}%
	\acro{BI}{Beacon Interval}%
	\acro{BLE}{Bluetooth Low Energy}%
	\acro{BO}{Beacon Order}%
	\acro{BPSK}{Binary Phase Shift Keying}%
	\acro{BSD}{Berkeley Software Distribution}%
	\acro{BSR}{Buffer Status Report}%
	\acro{BSS}{Basic Service Set}%
	\acro{BTP}{Basic Transport Protocol}%
	\acro{CA}{Cooperative Awareness}%
	\acro{CACC}{Cooperative Adaptive Cruise Control}%
	\acro{CAM}{\ac{CA} Message}
	\acro{CAN}{Controller Area Network}%
	\acro{CAP}{Contention Access Period}%
	\acro{CBR}{\todo{ambiguous acronym -- need to modify the multiple meanings}}%
	\acro{CoBR}{Constant Bit Rate}%
	\acro{ChBR}{Channel Busy Ratio}%
	\acro{CC}{Cruise Control}%
	\acro{CCA}{Clear Channel Assessment}%
	\acro{CDAP}{Common Distributed Application Protocol}%
	\acro{CEP-id}{Connection Endpoint \acl{ID}}%
	\acro{CFP}{Contention Free Period}%
	\acro{CLI}{Command Line Interface}%
	\acro{CoAP}{Constrained Application Protocol}%
	\acro{CoMP}{Coordinated MultiPoint}%
	\acro{CoMP-CS}{\acl{CoMP} Coordinated Scheduling}%
	\acro{CoV}{Coefficient of Variation}%
	\acro{CQI}{Channel Quality Index}%
	\acro{CRC}{Cyclic Redundancy Check}%
	\acro{CS}{Convergence Sublayer}%
	\acro{CSMA}{Carrier Sense Multiple Access}%
	\acro{CSMA-CA}{\acl{CSMA} Collision Avoidance}%
	\acro{CSS}{Chirp Spread Spectrum}%
	\acro{CSV}{Comma Separated Values}%
	\acro{CTS}{Clear to Send}%
	\acro{C-V2X}{Cellular \ac{V2X}}%
	\acro{CW}{Contention Window Length}%
	\acro{D2D}{Device-to-Device}%
	\acro{DA}{\acl{DIF} Allocator}%
    \acro{DAD}{Duplicate Address Detection}%
	\acro{DAF}{Distributed Application Facility}%
	\acro{DAN}{Distributed Application Name}%
	\acro{DAP}{Distributed Application Process}%
	\acro{DCC}{Decentralized Congestion Control}%
	\acro{DEN}{Decentralized Environmental Notification}%
	\acro{DENM}{\ac{DEN} Message}%
	\acro{DES}{Discrete Event Simulation}%
	\acro{DDS}{Data Distribution Service}%
	\acro{DGA}{Deterministic Gossip Algorithm}%
	\acro{DHCP}{Dynamic Host Configuration Protocol}%
	\acro{DIF}{Distributed \acl{IPC} Facility}%
	\acro{DL}{Downlink}%
	\acro{DLL}{Data Link Layer}%
	\acro{DNS}{Domain Name System}%
	\acro{DOI}{Digital Object Identifier}%
	\acro{DRF}{Data Run Flag}%
	\acro{DSCP}{Differentiated Services Control Point}%
	\acro{DSDV}{Destination-Sequenced Distance Vector}%
	\acro{DSL}{Domain Specific Language}%
    \acro{DSME}{Deterministic and Synchronous Multi-Channel Extension}%
	\acro{DSR}{Dynamic Source Routing}%
	\acro{DSRC}{Dedicated Short Range Communication}%
	\acro{DSSS}{Direct Sequence Spread Spectrum}%
	\acro{DTCP}{Data Transfer Control Protocol}%
	\acro{DTLS}{Datagram Transport Layer Security}%
	\acro{DTN}{Delay Tolerant Network}%
	\acro{DTP}{Data Transfer Protocol}%
	\acro{DUT}{Device-under-Test}%
	\acro{DYMO}{Dynamic \acl{MANET} On-demand}%
	\acro{E2E}{End-to-End}%
	\acro{ECN}{Explicit Congestion Notification}%
	\acro{ECU}{Electronic Control Unit}%
	\acro{ED}{Energy Detection}%
	\acro{EFCP}{Error and Flow Control Protocol}%
    \acro{EFCPI}{\acl{EFCP} Instance}%
	\acro{EIRP}{Equivalent Isotropically Radiated Power}%
	\acro{EMC}{Electromagnetic Compatibility}%
	\acro{eNB}{evolved Node B}%
	\acro{EPC}{Evolved Packet Core}%
	\acro{ER}{Edge Router}%
	\acro{ESP}{Encapsulating Security Payload}%
	\acro{ETSI}{European Telecommunications Standards Institute}%
	\acro{EUI}{Extended Unique Identifier}%
	\acro{EUI-64}{64-bit Global Identifier}%
	\acro{EXI}{Efficient \acl{XML} Exchange}%
	\acro{FA}{Flow Allocator}%
    \acro{FAI}{\acl{FA} Instance}%
	\acro{FCS}{Frame Check Sequence}%
	\acro{FDD}{Frequency Division Duplex}%
	\acro{FDMA}{Frequency Division Multiple Access}%
	\acro{FEL}{Future Event List}%
	\acro{FFD}{Full-Function Device}%
	\acro{FIB}{Forwarding Information Base}%
	\acro{FIFO}{First In First Out}%
	\acro{FLoRa}{Framework for \acl{LoRa}}%
	\acro{FOT}{Field Operational Test}%
	\acro{FoV}{Field of View}%
	\acro{FP}{Frame Pending}%
	\acro{FPGA}{Field Programmable Gate Array}%
	\acro{FRH}{Frame Header}%
	\acro{FSK}{Frequency Shift Keying}%
	\acro{GCC}{GNU Compiler Collection}%
	\acro{GEM}{Global Environment Model}%
	\acro{GEP}{Global Event Processor}%
	\acro{GN}{GeoNetworking}%
	\acro{GPIO}{General Purpose Input Output}%
	\acro{GPL}{GNU General Public License}%
	\acro{GPRS}{General Packet Radio System}%
	\acro{GPS}{Global Positioning System}%
	\acro{GPSR}{Greedy Perimeter Stateless Routing}%
	\acro{GSM}{Global System for Mobile Communications}%
	\acro{GTS}{Guaranteed Time Slot}%
	\acro{GUI}{Graphical User Interface}%
	\acro{H2H}{Human-to-Human}%
	\acro{H-ARQ}{Hybrid Automatic Repeat reQuest}%
	\acro{HC}{Header Compression}%
	\acro{HIL}{Hardware-in-the-Loop}%
	\acro{HSS}{Home Subscriber Server}%
	\acro{HTML}{Hypertext Markup Language}%
	\acro{HTTP}{Hypertext Transfer Protocol}%
	\acro{HWMP}{Hybrid Wireless Mesh Protocol}%
	\acro{IANA}{Internet Assigned Numbers Authority}%
	\acro{IAS}{Intersection Assistance System}\acroplural{IAS}[IAS]{Intersection Assistance Systems}%
	\acro{ICMP}{Internet Control Message Protocol}%
	\acro{ICMPv6}{\acl{ICMP} for \acl{IPv6}}%
	\acro{ICT}{Information and Communication Technology}%
	\acro{ID}{Identifier}%
	\acro{IDE}{Integrated Development Environment}%
	\acro{IDM}{Intelligent Driver Model}%
	\acro{IEEE}{Institute of Electrical and Electronics Engineers}%
	\acro{IETF}{Internet Engineering Task Force}%
	\acro{IFQ}{Interface Queue}%
	\acro{IFS}{Inter Frame Spacing}%
	\acro{IID}{Interface Identifier}%
	\acro{IKEv2}{Internet Key Exchange Version 2}%
	\acro{IoT}{Internet of Things}%
	\acro{IP}{Internet Protocol}%
	\acro{IPC}{Inter-Process Communication}%
	\acro{IPCP}{\acl{IPC} Process}%
	\acro{IPsec}{Internet Protocol security}%
	\acro{IPSO}{IP for Smart Objects}%
	\acro{IPv4}{\acl{IP} Version 4}%
	\acro{IPv6}{\acl{IP} Version 6}%
	\acro{IrDA}{Infrared Data Association}%
	\acro{IRM}{\acl{IPC} Resource Manager}%
	\acro{ISA}{International Society for Automation}%
	\acro{ISM}{Industrial, Scientific and Medical}%
	\acro{ISO}{International Organization for Standardization}%
	\acro{ITS}{Intelligent Transportation System}\acroplural{ITS}[ITS]{Intelligent Transportation Systems}%
	\acro{ITS-S}{\acl{ITS}-Station}%
	\acro{IVC}{Inter-Vehicle Communication}%
	\acro{JNI}{Java Native Interface}%
	\acro{LAN}{Local Area Network}%
	\acro{LBR}{\acl{LLN} Border Router}%
	\acro{LCID}{Logical Connection \acl{ID}}%
	\acro{LDM}{Local Dynamic Map}%
	\acro{LDP}{Label Distribution Protocol}%
	\acro{LEM}{Local Environment Model}%
	\acro{LEP}{Local Event Processor}%
	\acro{LIMoSim}{Lightweight \acl{ICT}-centric Mobility Simulation}%
	\acro{LIN}{Local Interconnect Network}%
	\acro{LLC}{Logical Link Control}%
    \acro{LLDN}{Low Latency Deterministic Network}%
	\acro{LLN}{Low-power and Lossy Network}%
	\acro{LOS}{Line of Sight}%
	\acro{LoRa}{Long Range}%
	\acro{LoRaWAN}{\acl{LoRa} \acl{WAN}}%
	\acro{LoWPAN}{Low-power \acl{WPAN}}%
	\acro{LPDU}{\acl{LLC} \acl{PDU}}%
	\acro{LQI}{Link Quality Indicator}%
	\acro{LTE}{Long Term Evolution}%
	\acro{LTE-A}{\acl{LTE} Advanced}%
	\acro{LWM2M}{Lightweight \acl{M2M}}%
	\acro{M2M}{Machine-to-Machine}%
	\acro{MAC}{Medium Access Control}%
	\acro{MANET}{Mobile Ad Hoc Network}%
	\acro{MATSim-T}{Multi Agent Transport Simulation Toolkit}%
	\acro{MCPS}{\acl{MAC} Common Part Sublayer}%
	\acro{MCS}{Modulation and Coding Scheme}%
	\acro{MEC}{Multi-access Edge Computing}%
	\acro{MIB}{Management Information Base}%
	\acro{MIC}{Message Integrity Code}%
	\acro{MIMO}{Multiple Input Multiple Output}%
	\acro{MIP}{Mobile \acl{IP}}%
	\acro{MIPv6}{\acl{MIP} version 6}%
	\acro{MLME}{\acl{MAC} Sublayer Management Entity}%
	\acro{MME}{Mobile Management Entity}%
	\acro{mmWave}{millimeter Wave}%
	\acro{MOBIL}{Minimizing Overall Braking Induced by Lane change}%
	\acro{MPDU}{\acl{MAC} \acl{PDU}}%
	\acro{MPEG}{Motion Picture Experts Group}%
	\acro{MPEG-4}{\acl{MPEG} Layer-4 Video}%
	\acro{MPLS}{Multiprotocol Label Switching}%
	\acro{MPP}{Mobile \acl{P2P} Protocol}%
	\acro{MO}{Multi-Superframe Order}%
	\acro{MOST}{Media Oriented Systems Transport}%
	\acro{MQTT}{Message Queuing Telemetry Transport}%
	\acro{MQTT-S}{Message Queuing Telemetry Transport for Sensor Networks}%
	\acro{MSB}{Most Significant Bit}%
	\acro{MSDU}{\acl{MAC} \acl{SDU}}%
	\acro{MTC}{Machine Type Communication}%
	\acro{MTU}{Maximum Transfer Unit}%
	\acro{NA}{Neighbor Advertisement}%
	\acro{NAT}{Network Address Translation}%
	\acro{NB}{Number of Backoffs}%
	\acro{NC}{Node Confirmation}%
	\acro{ND}{Neighbor Discovery}%
	\acro{NED}{Network Topology Description}%
	\acro{NETA}{NETwork Attacks Framework for \omnet{}}%
	\acro{NIC}{Network Interface Card}%
	\acro{NLOS}{Non Line of Sight}%
	\acro{NR}{Node Registration}%
	\acro{ns-2}{Network Simulator 2}%
	\acro{ns-3}{Network Simulator 3}%
	\acro{NS}{Neighbor Solicitation}%
	\acro{NSM}{Namespace Management}%
	\acro{OASIS}{Organization for the Advancement of Structured Information Standards}%
	\acro{OCB}{Outside the Context of a \acl{BSS}}%
	\acro{ODD}{Organic Data Dissemination}%
	\acro{OFDM}{Orthogonal Frequency Division Multiplex}%
	\acro{OFDMA}{Orthogonal Frequency Division Multiple Access}%
	\acro{OGC}{Open Geospatial Consortium}%
	\acro{OLSR}{Optimized Link State Routing}%
	\acro{ONE}{Opportunistic Network Environment}%
	\acro{OPEN}{One-Pair Ether-Net}%
	\acro{OpenGL}{Open Graphics Library}%
	\acro{OppNet}{Opportunistic Network}%
    \acro{OPS}{Opportunistic Protocol Simulator}%
	\acro{O-QPSK}{Offset Quadrature Phase Shift Keying}%
	\acro{OSCORE}{Object Security for Constrained RESTful Environments}%
	\acro{OSI}{Open Systems Interconnection}%
	\acro{OSM}{OpenStreetMap}%
	\acro{OSPF}{Open Shortest Path First}%
	\acro{P2P}{Peer-to-Peer}%
	\acro{PAN}{Personal Area Network}%
	\acro{PAN-ID}{\acl{PAN} Identifier}%
	\acro{PASER}{Position Aware Secure and Efficient Mesh Routing Protocol}%
	\acro{PCAP}{Packet Capture}%
	\acro{PD}{\acl{PHY} Data Service}%
	\acro{PDA}{Personal Digital Assistant}%
	\acro{PDCP}{Packet Data Convergence Protocol}%
	\acro{PDR}{Packet Delivery Ratio}%
	\acro{PDU}{Protocol Data Unit}%
	\acro{PDUFG}{\acl{PDU} Forwarding Generator}%
	\acro{PER}{Packet Error Rate}%
	\acro{PFH}{\acl{PCAP} File Header}%
	\acro{PGW}{Packet Data Network Gateway}%
	\acro{PHR}{\acl{PHY} Header}%
	\acro{PHY}{Physical Layer}%
	\acro{PIB}{\acl{PAN} Information Base}%
	\acro{PLC}{Power Line Communication}%
	\acro{PLME}{\acl{PHY} Management Entity}%
	\acro{PM}{\acl{PHY} Management Service}%
	\acro{POS}{Personal Operating Space}%
	\acro{PPDU}{\acl{PHY} \acl{PDU}}%
	\acro{PPP}{Point-to-Point Protocol}%
	\acro{PSDU}{\acl{PHY} \acl{SDU}}%
	\acro{PSSS}{Parallel Sequence Spread Spectrum}%
	\acro{QAM}{Quadrature Amplitude Modulation}%
	\acro{QoS}{Quality of Service}%
	\acro{QPSK}{Quadrature Phase Shift Keying}%
	\acro{RA}{\todo{double meaning -- need to modify this}}%
	\acro{RAd}{Router Advertisement}%
	\acro{RAl}{Resource Allocator}%
	\acro{RAC}{Random Access}%
	\acro{RAM}{Random Access Memory}%
	\acro{RAN}{Radio Access Network}%
	\acro{RB}{Resource Block}%
	\acro{RDP}{Remote Desktop Protocol}%
	\acro{REST}{Representational State Transfer}%
	\acro{RF}{Radio Frequency}%
	\acro{RFC}{Request for Comments}%
	\acro{RFD}{Reduced-Function Device}%
	\acro{RFID}{Radio-Frequency Identification}%
	\acro{RIB}{Resource Information Base}%
    \acro{RIBd}{\acl{RIB} Daemon}%
	\acro{RINA}{Recursive InterNetwork Architecture}%
	\acro{RLC}{Radio Link Control}%
	\acro{RLWE}{Receiver's Left Window Edge}%
	\acro{RMT}{Relaying and Multiplexing Task}%
	\acro{ROHC}{RObust Header Compression}%
	\acro{ROI}{Region of Interest}%
	\acro{ROM}{Read-Only Memory}%
	\acro{ROLL}{Routing over Low-power and Lossy networks}%
	\acro{ROS}{Robot Operating System}%
	\acro{RPC}{Remote Procedure Call}%
	\acro{RPL}{Routing Protocol for Low power and Lossy Networks}%
	\acro{RREP}{Route Reply}%
	\acro{RREQ}{Route Request}%
	\acro{RRS}{Randomized Rumor Spreading}%
	\acro{RRWE}{Receiver's Right Window Edge}%
	\acro{RS}{Router Solicitation}%
	\acro{RSU}{Roadside Unit}%
	\acro{RSS}{Received Signal Strength}%
	\acro{RSSI}{Received Signal Strength Indicator}%
	\acro{RSTP}{Rapid Spanning Tree Protocol}%
	\acro{RSVP}{Resource Reservation Protocol}%
	\acro{RSVP-TE}{\acl{RSVP} - Traffic Engineering}%
	\acro{RTCP}{\acl{RTP} Control Protocol}%
	\acro{RTP}{Real-time Transport Protocol}%
	\acro{RTS}{Request to Send}%
	\acro{RTT}{Round-Trip Time}%
	\acro{RX}{Receiving}%
	\acro{SAA}{Stateless Address Auto-configuration}%
	\acro{SAP}{Service Access Point}%
	\acro{SAB}{Slot Allocation Bitmap}%
	\acro{SAS}{Smart Antenna System}%
	\acro{SCADA}{Supervisory Control and Data Acquisition}%
	\acro{SCF}{Store \& Carry Forwarding}%
	\acro{SCTP}{Stream Control Transmission Protocol}%
	\acro{SSCS}{Service Specific Convergence Sublayer}%
	\acro{SD}{Superframe Duration}%
	\acro{SDN}{Software Defined Networking}%
	\acro{SDU}{Service Data Unit}%
	\acro{SFD}{Start of Frame Delimiter}%
	\acro{SGW}{Serving Gateway}%
	\acro{SHB}{Single Hop Broadcast}%
	\acro{SHR}{Synchronization Header}%
	\acro{SI}{Superframe Interval}%
	\acro{SICS}{Swedish Institute of Computer Science}%
	\acro{SINR}{Signal-to-Interference-plus-Noise-Ratio}%
	\acro{SIP}{Session Initiation Protocol}%
	\acro{SLWE}{Sender's Left Window Edge}%
	\acro{SNMP}{Simple Network Management Protocol}%
	\acro{SNR}{Signal-to-Noise-Ratio}%
	\acro{SO}{Superframe Order}%
	\acro{SOAP}{Simple Object Access Protocol}%
	\acro{SON}{Self Organizing Network}%
	\acro{SPI}{Serial Peripheral Interface}%
	\acro{SRWE}{Sender's Right Window Edge}%
    \acro{STL}{Standard Template Library}
	\acro{SUMO}{Simulation of Urban MObility}%
	\acro{SV}{Summary Vector}%
	\acro{SWIM}{Small Worlds in Motion}%
	\acro{TB}{Transport Block}%
	\acro{TCP}{Transmission Control Protocol}%
	\acro{TDD}[\todo{FIXME/TODO: DO NOT USE THIS ACRONYM - IT IS AMBIGUOUS}]{\todo{FIXME/TODO: DO NOT USE THIS ACRONYM - IT IS AMBIGUOUS}}%
	\acro{TDDtime}[TDD]{Time Division Duplex}%
	\acro{TDDtest}[TDD]{Test Driven Development}%
	\acro{TDMA}{Time Division Multiple Access}%
	\acro{TI}{Texas Instruments}%
	\acro{TLS}{Transport Layer Security}%
	\acro{TPC}{Transmission Power Control}%
	\acro{TraCI}{Traffic Control Interface}%
	\acro{TSCH}{Time Slotted Channel Hopping}%
	\acro{TSN}{Time-Sensitive Networking}%
	\acro{TTC}{Time To Collision}%
	\acro{TTI}{Transmission Time Interval}%
	\acro{TTL}{Time To Live}%
	\acro{TX}{Transmitting}%
	\acro{TXOP}{Transmit Opportunity}%
	\acro{UART}{Universal Asynchronous Receiver and Transmitter}%
	\acro{UAV}{Unmanned Aerial Vehicle}%
	\acro{USART}{Universal Synchronous Asynchronous Receiver and Transmitter}%
	\acro{UDG}{Unit Disk Graph}%
	\acro{UDP}{User Datagram Protocol}%
	\acro{UE}{User Equipment}%
	\acro{UHF}{Ultra High Frequency}%
	\acro{UI}{User Interface}%
	\acro{UL}{Uplink}%
	\acro{ULA}{Uniform Linear Array}%
	\acro{UML}{Unified Modeling Language}%
	\acro{UMTS}{Universal Mobile Telecommunications System}%
	\acro{URI}{Uniform Resource Identifier}%
	\acro{URL}{Uniform Resource Locator}%
	\acro{URLLC}{Ultra-reliable Low Latency Communication}%
	\acro{USB}{Universal Serial Bus}%
	\acro{UWB}{Ultra-Wideband}%
	\acro{V2I}{Vehicle-to-Infrastructure}%
	\acro{V2V}{Vehicle-to-Vehicle}%
	\acro{V2X}{Vehicle-to-Everything}%
	\acro{VANET}{Vehicular Ad Hoc Network}%
	\acro{VDP}{Vehicle Data Provider}%
	\acro{Veins}{Vehicles in Network Simulation}%
	\acro{VHT}{Very High Throughput}%
	\acro{VLC}{Visible Light Communication}%
	\acro{VoD}{Video on Demand}%
	\acro{VoIP}{Voice-over-IP}%
	\acro{VTB}{Virtual Testbed}%
	\acro{VPN}{Virtual Private Network}%
	\acro{W3C}{World Wide Web Consortium}%
	\acro{WAMP}{Web Application Messaging Protocol}%
	\acro{WAN}{Wide Area Network}%
	\acro{WAVE}{Wireless Access in Vehicular Environments}%
	\acro{WBAN}{Wireless Body Area Network}%
	\acro{W-CDMA}{Wideband Code Division Multiple Access}%
	\acro{WCPS}{Wireless Cyber-Physical System}%
	\acro{WGS84}{World Geodetic System 1984}%
	\acro{WiMAX}{Worldwide Interoperability for Microwave Access}%
	\acro{WLAN}{Wireless Local Area Network}%
	\acro{WPAN}{Wireless Personal Area Network}%
	\acro{WSAN}{Wireless Sensor and Actuator Network}%
	\acro{WSA}{Wave Service Advertisement}%
	\acro{WSM}{Wave Short Message}%
	\acro{WSN}{Wireless Sensor Network}%
	\acro{WWW}{World Wide Web}%
	\acro{XEP}{\acl{XMPP} extension}%
	\acro{XML}{Extensible Markup Language}%
	\acro{XMPP}{Extensible Messaging and Presence Protocol}%
	\acro{XSD}{\acl{XML} Schema Definition}%
	\acro{ZAL}{ZigBee Application Layer}%
	\acro{ZC}{ZigBee Coordinator}%
	\acro{ZCL}{ZigBee Cluster Library}%
	\acro{ZDO}{ZigBee Device Object}%
	\acro{ZED}{ZigBee End Device}%
	\acro{ZR}{ZigBee Router}%
\end{acronym}
\end{multicols}
%